\documentclass[11pt]{article}
\usepackage{amssymb}
\usepackage{graphics}
\usepackage{epsfig}
\usepackage{a4wide}
\usepackage{color}
\usepackage{multirow}
\usepackage{cite}

\begin{document}

\title{ Dimension-six operators in Higgs pair production via vector-boson fusion at the LHC }
\author{ Ling Liu-Sheng$^1$, Zhang Ren-You$^1$, Ma Wen-Gan$^1$, Li Xiao-Zhou$^1$, Guo Lei$^2$ and Wang Shao-Ming$^2$ \\
{\small $^1$ Department of Modern Physics, University of Science and Technology of China (USTC),}  \\
{\small  Hefei 230026, Anhui, People's Republic of China} \\
{\small $^2$ Department of Physics, Chongqing University, Chongqing 401331, People's Republic of China}  }
\date{}
\maketitle \vskip 15mm

\begin{abstract}
The effective Lagrangian formalism provides a way to study the new physics effects at the electroweak scale. We study the Higgs pair production via the vector-boson fusion (VBF) at the Large Hadron Collider within the framework of the effective field theory. The effects from the dimension-six operators involved in the VBF Higgs pair production, particularly $\mathcal{O}_{\Phi,2}$ and $\mathcal{O}_{\Phi,3}$ which are relevant to the triple Higgs self-coupling, on the integrated cross section and various kinematic distributions are investigated. We find that the distributions of Higgs pair invariant mass, Higgs transverse momentum and rapidity are significantly altered by the operators $\mathcal{O}_{\Phi,2}$ and $\mathcal{O}_{\Phi,3}$. These features are helpful in disentangling the contributions from the operators $\mathcal{O}_{\Phi,2}$ and $\mathcal{O}_{\Phi,3}$ in triple Higgs self-coupling. We also provide the 5$\sigma$ discovery and 3$\sigma$ exclusion limits for the coefficients of $\mathcal{O}_{\Phi,2}$ and $\mathcal{O}_{\Phi,3}$ by measuring the VBF Higgs pair production process including the sequential $H \to b\bar{b}$ decays at the $14~ {\rm TeV}$ LHC.
\end{abstract}
\vskip 3cm


\vfill \eject
\baselineskip=0.32in

\vskip 5mm
\section{Introduction} \label{section-I}
\par
After the discovery of the $125~{\rm GeV}$ Higgs boson, the next important step is to perform detailed study of this new boson \cite{hd-1,hd-2}. The Higgs self-couplings are extremely significant for understanding the electroweak symmetry breaking (EWSB). The Standard Model (SM) contains triple and quartic Higgs self-couplings. The measurement of the quartic Higgs self-coupling is a serious challenge even at the foreseen collider, such as the Future Circular Collider in hadron-hadron mode (FCC-hh) with colliding energy of $100~{\rm TeV}$ \cite{collider-2,self-1}, while the triple Higgs self-coupling is accessible through the Higgs boson pair production at the Large Hadron Collider (LHC) and the FCC-hh \cite{self-2,self-3}. The searches for Higgs boson pair production performed by the ATLAS and CMS collaborations have shown that the upper limit of its cross section is corresponding to dozens times of the SM prediction \cite{selfdect}, and there is still room left for a beyond Standard Model (BSM) interpretation.

\par
Since we do not know the specific form of the BSM theory and no new particle BSM has been observed, we can investigate the new physics by adopting the effective field theory (EFT) approach \cite{eft-1}. In an EFT the SM is considered as an effective low-energy theory, while the higher dimensional interaction terms appear in the Lagrangian providing an adequate low-energy description for the physics BSM. Actually, the effects from many new physics models arise to the first approximation at the level of dimension-six operators \cite{Henning}, and that makes it a very popular EFT approach for LHC searches of BSM.

\par
In the SM the gluon-gluon fusion (GGF) mechanism provides the largest cross section contribution for the Higgs pair production \cite{hpair-tot}. The effects of dimension-six operators on GGF Higgs pair production have been studied in Refs.\cite{hpair-1,hpair-2,hpair-3,hpair-4,hpair-5,hpair-6,hpair-7,hpair-8,hpair-9}. The vector-boson fusion (VBF) Higgs pair production mechanism yields the second largest cross section and offers a clean experimental signature of two centrally produced Higgs bosons with two hard jets in the forward/backward rapidity region \cite{vbfsig}. In this work we investigate the details of the VBF Higgs pair production at the $\sqrt{S}=13,~14~{\rm TeV}$ LHC so as to derive information regarding the dimension-six operators relevant to Higgs couplings. The paper is organized as follows: In Section 2, we introduce the dimension-six operators concerned in VBF Higgs pair production at the LHC. Section 3 presents the calculation and analysis strategies. The numerical results and discussion are illustrated in Section 4. Finally a short summary is given in Section 5.

\vskip 5mm
\section{Theoretical Framework}
\label{section-II}
\par
Assuming baryon and lepton number conservation, the full EFT Lagrangian takes the form as \cite{eft-1,eft-2,Andre}
\begin{eqnarray}
\label{Lagrange}
\mathcal{L}_{\rm eff}= \mathcal{L}_{\rm SM}
                     + \sum_{i}\frac{f^{(6)}_i}{\Lambda^2}\mathcal{O}^{(6)}_i
                     + \sum_{i}\frac{f^{(8)}_i}{\Lambda^4}\mathcal{O}^{(8)}_i + \cdots \, ,
\end{eqnarray}
where $\Lambda$ is the large scale of a given BSM theory, the experimental bounds on neutrino masses indicate a strong suppression by $\Lambda \sim 10^{14}~{\rm GeV}$ \cite{Ben}. $\mathcal{L}_{\rm SM}$ is the SM Lagrangian, $\mathcal{O}^{(n)}_i~(n=6,8,...)$ stand for the dimension-$n$ operators that respect the SM gauge invariance, and $f^{(n)}_i$ are dimensionless running as functions $f^{(n)}_i(\mu)$ of the renormalization group (RG) scale $\mu$ \cite{eft-2,Grogean}. The nonrenormalizable higher dimensional operators induce the partial-wave unitarity violation by leading to rapid growth of the scattering amplitudes with the energy. Once the coefficients $f^{(n)}_i$ are fixed, this fact constrains the energy range where the low-energy EFT is valid \cite{eft-5}. In this EFT the lowest order operators that describe the new physics are dimension-six operators.

\par
In the LHC physics at the present phase, the operators with dimension larger than six are generally only sub-leading with respect to the dimension-six operators and can be neglected, excepting some cases with special symmetry constructions \cite{Roberto}. For simplicity, as a first step we adopt the single dominance hypothesis that all the anomalous couplings except the dimension-six operators are zero in this work. This assumption might be valid in a variety of BSM theories, e.g., in strongly interacting BSM theories, the pure dimension-eight operators at the amplitude level will interfere with the weakly SM amplitude at the cross section level, thus, giving a natural suppression in comparison with the strongly $A_{{\mathcal{O}}^{(6)}} \times A_{{\mathcal{O}}^{(6)}}$ contributions, and the technically contributions of dimension eight at the cross section level due to ${\mathcal{O}}^{(6)} \times {\mathcal{O}}^{(6)}$ contributions can be neglected.

\par
The tree-level VBF Higgs pair production at the LHC involves the fermion gauge couplings, Higgs gauge couplings and triple Higgs self-coupling. Since the fermion gauge couplings are in agreement with the SM predictions at the per mil level and the operators that modify these couplings are severely constrained \cite{eft-3,constraint-1}, we neglect the new physics effect on the fermion gauge couplings. Considering the $C$ and $P$ conservation, there are nine dimension-six operators which modify the Higgs gauge couplings and Higgs self-couplings. We expressed these operators in the forms as \cite{eft-5}
\begin{eqnarray}
\label{operators}
&&
{\cal O}_{\Phi,1}= (D_{\mu}\Phi)^{\dag}\Phi\Phi^{\dag}(D^{\mu}\Phi),~~~
{\cal O}_{\Phi,2}= \frac{1}{2}\partial^{\mu}(\Phi^{\dag}\Phi)\partial_{\mu}(\Phi^{\dag}\Phi), \nonumber \\
&&
{\cal O}_{\Phi,3}= \frac{1}{3}(\Phi^{\dag}\Phi)^3,~~~~~~~~~~~~~~~\,
{\cal O}_{\Phi,4}=(D_\mu\Phi)^\dagger(D^\mu\Phi)(\Phi^\dagger\Phi), \nonumber \\
&&
{\cal O}_{B}= (D_{\mu}\Phi)^{\dag}\hat{B}^{\mu\nu}(D_{\nu}\Phi),~~~~~
{\cal O}_{W}= (D_{\mu}\Phi)^{\dag}\hat{W}^{\mu\nu}(D_{\nu}\Phi), \nonumber \\
&&
{\cal O}_{BB}= \Phi^{\dag}\hat{B}_{\mu\nu}\hat{B}^{\mu\nu}\Phi,~~~~~~~~~~\,
{\cal O}_{WW}= \Phi^{\dag}\hat{W}_{\mu\nu}\hat{W}^{\mu\nu}\Phi,~~~~~~~~~
{\cal O}_{BW}= \Phi^{\dag}\hat{B}_{\mu\nu}\hat{W}^{\mu\nu}\Phi,~~~~~~~~
\end{eqnarray}
where $\Phi$ stands for the SM Higgs doublet
\begin{eqnarray}
\Phi = {G^+ \choose \frac{v+H+iG^0}{\sqrt{2}}},
\end{eqnarray}
$G^+$ and $G^0$ are charged and neutral Goldstone bosons, respectively, $v$ denotes the vacuum expectation value (VEV), $H$ is the SM Higgs boson, $\hat{W}_{\mu \nu}=i\frac{g}{2}\sigma^a W_{\mu \nu}^a$ and $\hat{B}_{\mu \nu}=i\frac{g'}{2}B_{\mu \nu}$ with $g$ and $g^{\prime}$ being the $SU(2)_L$ and $U(1)_Y$ gauge couplings, respectively, and $\sigma^a~ (a=1,2,3)$ are the Pauli matrices.

\par
The dimension-six operators in Eq.(\ref{operators}) are classified into three sets,
\begin{eqnarray}
&&
A = \{ \mathcal{O}_{BW},~ \mathcal{O}_{\Phi,1} \} \nonumber \\
&&
B = \{ \mathcal{O}_{B},~ \mathcal{O}_{W},~ \mathcal{O}_{BB},~ \mathcal{O}_{WW} \} \nonumber \\
&&
C = \{ \mathcal{O}_{\Phi,2},~ \mathcal{O}_{\Phi,3},~ \mathcal{O}_{\Phi,4} \}\,.~~~~~~~
\end{eqnarray}
The operators in set $A$ (i.e., $\mathcal{O}_{BW}$ and $\mathcal{O}_{\Phi,1}$) contribute to the oblique electroweak precision parameters $S$ and $T$ at tree level \cite{st-1}. The measurements of $S$ and $T$ lead to stringent bounds on the two operators, and thus it is reassuring to neglect the effects from $\mathcal{O}_{BW}$ and $\mathcal{O}_{\Phi,1}$ \cite{eft-4}. The operators in set $B$ are related by two equations of motion of Higgs and electroweak gauge bosons \cite{eft-2, eft-3}, therefore only two of them are independent operators. The operators $\mathcal{O}_{\Phi,2}$, $\mathcal{O}_{\Phi,3}$ and $\mathcal{O}_{\Phi,4}$ in set $C$ are related to another dimension-six operator $\mathcal{O}_{\Phi,f}$ by an equation of motion of Higgs, electroweak gauge bosons and fermions \cite{eft-2,hpair-5}, where $\mathcal{O}_{\Phi,f}$ is given by
\begin{eqnarray}
\mathcal{O}_{\Phi,f}=(\Phi^\dagger \Phi)(\bar{f}_L \Phi f_R)+{\rm h.c.},
\end{eqnarray}
and $f_L$ and $f_R$ are $SU(2)_L$ doublet and singlet fermions, respectively. It implies that each operator in set $C$ can be expressed in terms of the other two operators in set $C$ and $\mathcal{O}_{\Phi,f}$. Based on above mentioned facts, only four independent dimension-six operators are considered in the VBF Higgs pair production, which are chosen as $\mathcal{O}_{BB}$, $\mathcal{O}_{WW}$, $\mathcal{O}_{\Phi,2}$ and $\mathcal{O}_{\Phi,3}$ in this paper.

\par
As shown in Eq.(\ref{operators}), $\mathcal{O}_{BB}$ and $\mathcal{O}_{WW}$ consist of two Higgs doublets and two gauge field strengths and thus affect the Higgs gauge couplings, while $\mathcal{O}_{\Phi,2}$ and $\mathcal{O}_{\Phi,3}$ consist of only Higgs doublets and modify only the Higgs self-couplings. In the following, we investigate the effects from these four operators on the VBF Higgs pair production at the LHC, particularly the latter two operators which modify the triple Higgs self-coupling. The related EFT Lagrangian up to dimension-six operators is written as
\begin{eqnarray}
\mathcal{L}_{\rm eff}= \mathcal{L}_{\rm SM} + \sum_{i} \mathcal{C}_{i} \, \mathcal{O}_{i},
\end{eqnarray}
where the coefficient $\mathcal{C}_{i}=f_i^{(6)}/\Lambda^2$ and $\mathcal{O}_{i}$ represents operator $\mathcal{O}_{BB}$, $\mathcal{O}_{WW}$, $\mathcal{O}_{\Phi,2}$ or $\mathcal{O}_{\Phi,3}$. The operator $\mathcal{O}_{\Phi,2}$ modifies the Higgs kinetic term as
\begin{eqnarray}
\frac{1}{2} \left(1+ v^2 \mathcal{C}_{\Phi,2} \right) \partial^\mu H \partial_\mu H.
\end{eqnarray}
We redefine Higgs boson field as $\left( 1+v^2 \mathcal{C}_{\Phi,2} \right)^{\frac{1}{2}}H$ to obtain canonical form of the Higgs kinetic term. This redefinition shifts all Higgs couplings with SM particles and Higgs boson mass. The effective Higgs potential is given by
\begin{eqnarray}
&&
V_{\rm eff}(\Phi)
=
V_{\rm SM}(\Phi) - \frac{1}{3} \mathcal{C}_{\Phi,3} (\Phi^{\dagger} \Phi)^3 \nonumber \\
&&~~~~~~~~~
=
-\mu^2 \Phi^{\dagger} \Phi + \frac{\lambda}{4} (\Phi^{\dagger} \Phi)^2
          -\frac{1}{3} \mathcal{C}_{\Phi,3} (\Phi^{\dagger} \Phi)^3,~~~
\end{eqnarray}
where the additional term $-\frac{1}{3} \mathcal{C}_{\Phi,3} (\Phi^{\dagger} \Phi)^3$ is induced by the operator $\mathcal{O}_{\Phi,3}$. This additional term modifies the minimum of the Higgs potential as well as the relation between $\lambda$, $v$ and Higgs mass $M_H$. Consequently, up to the linear order in the coefficients $\mathcal{C}_{\Phi,2}$ and $\mathcal{C}_{\Phi,3}$, the Higgs boson mass reads
\begin{eqnarray}
\label{mhvl}
M_H^2=\frac{1}{2} \lambda v^2
\left(
1-v^2 \mathcal{C}_{\Phi,2}-\frac{2}{\lambda} v^2 \mathcal{C}_{\Phi,3}
\right).
\end{eqnarray}

\par
For the VBF Higgs pair production at a proton-proton collider, the operators $\mathcal{O}_{BB}$, $\mathcal{O}_{WW}$ and $\mathcal{O}_{\Phi,2}$ shift the $HVV$ and $HHVV$ Higgs gauge couplings, and both $\mathcal{O}_{\Phi,2}$ and $\mathcal{O}_{\Phi,3}$ modify the $HHH$ Higgs self-coupling. The redefinition of the Higgs field induced by $\mathcal{O}_{\Phi,2}$ contributes a factor $\left( 1-\frac{v^2}{2} \mathcal{C}_{\Phi,2} \right)$ to the $HGV$ Higgs-Goldstone-gauge-boson couplings. The effective Lagrangian for $HVV$ and $HHVV$ interactions are written as \cite{eft-5}
\begin{eqnarray}
&&~~
\mathcal{L}^{HVV}_{\rm eff}
=
g_{HAA}HA_{\mu\nu}A^{\mu\nu} + g_{HZA}HA_{\mu\nu}Z^{\mu\nu} + g^{(1)}_{HZZ}HZ_{\mu\nu}Z^{\mu\nu}+g^{(2)}_{HZZ}HZ_{\mu}Z^{\mu} ~~~~ \nonumber \\
&&~~~~~~~~~~~~~\,
+ g^{(1)}_{HWW}HW^+_{\mu\nu}W^{-\mu\nu} +g^{(2)}_{HWW}HW^+_{\mu}W^{-\mu}, \nonumber \\
&&
\mathcal{L}^{HHVV}_{\rm eff}
=
g_{HHAA}H^2A_{\mu\nu}A^{\mu\nu} + g_{HHZA}H^2A_{\mu\nu}Z^{\mu\nu} + g^{(1)}_{HHZZ}H^2Z_{\mu\nu}Z^{\mu\nu} \nonumber \\
&&~~~~~~~~~~~~~\,
+g^{(2)}_{HHZZ}H^2Z_{\mu}Z^{\mu} + g^{(1)}_{HHWW}H^2W^+_{\mu\nu}W^{-\mu\nu} + g^{(2)}_{HHWW}H^2W^+_{\mu}W^{-\mu},
\end{eqnarray}
with $V_{\mu\nu}=\partial_{\mu}V_{\nu}-\partial_{\nu}V_{\mu}~(V=A,Z,W^{\pm})$. The corresponding coupling constants are given by
\begin{eqnarray}
\label{Rule-vertex-1}
&&
g_{HAA} = -\frac{s_W^2}{4} g^2 v \Big( \mathcal{C}_{BB} + \mathcal{C}_{WW} \Big),~~~~~~~~~~~~~
g_{HZA} = \frac{s_W}{2 c_W} g^2v \Big( s_W^2 \mathcal{C}_{BB} - c^2_W\mathcal{C}_{WW} \Big),~~~~~~ \nonumber  \\
&&
g^{(1)}_{HZZ} = -\frac{1}{4 c_W^2} g^2v \Big( s^4_W\mathcal{C}_{BB}+c^4_W\mathcal{C}_{WW} \Big),~~~\,
g^{(2)}_{HZZ} = \frac{1}{4c^2_W} g^2v \Big( 1-\frac{v^2}{2}\mathcal{C}_{\phi,2} \Big), \nonumber  \\
&&
g^{(1)}_{HWW} = -\frac{1}{2} g^2v \, \mathcal{C}_{WW}, ~~~~~~~~~~~~~~~~~~~~~~~~~~
g^{(2)}_{HWW} = \frac{1}{2} g^2v \Big( 1-\frac{v^2}{2}\mathcal{C}_{\phi,2} \Big),
\end{eqnarray}
and
\begin{eqnarray}
\label{Rule-vertex-2}
&&
g_{HHAA} = -\frac{s^2_W}{8} g^2 \Big( \mathcal{C}_{WW} + \mathcal{C}_{BB} \Big),~~~~~~~~~~~~~~
g_{HHZA} = -\frac{s_W}{4c_W} g^2 \Big( c^2_W \mathcal{C}_{WW} - s^2_W \mathcal{C}_{BB} \Big),~~~~ \nonumber  \\
&&
g^{(1)}_{HHZZ} = -\frac{1}{8c^2_W} g^2 \Big( c^4_W\mathcal{C}_{WW} + s^4_W \mathcal{C}_{BB} \Big),~~~~\,
g^{(2)}_{HHZZ} = \frac{1}{8c^2_W} g^2 \Big( 1 - v^2\mathcal{C}_{\phi,2} \Big), \nonumber  \\
&&
g^{(1)}_{HHWW} = -\frac{1}{4} g^2 \, \mathcal{C}_{WW},~~~~~~~~~~~~~~~~~~~~~~~~~~~
g^{(2)}_{HHWW} = \frac{1}{4} g^2 \Big( 1 - v^2 \mathcal{C}_{\phi,2} \Big),
\end{eqnarray}
where $s_W = \sin\theta_W$, $c_W = \cos\theta_W$, $\theta_W$ is the weak mixing angle, $g = e/\sin\theta_W$, and $e$ represents the electric charge of positron. The effective Lagrangian for triple Higgs self-interaction has the form as
\begin{eqnarray}
\label{l-vertex-3}
\mathcal{L}^{HHH}_{\rm eff} = g^{(1)}_{HHH}H^3+g^{(2)}_{HHH}H(\partial_{\mu}H)(\partial^{\mu}H),
\end{eqnarray}
with
\begin{eqnarray}
\label{Rule-vertex-3}
g^{(1)}_{HHH} = -\frac{\lambda}{4}v +\frac{3 \lambda}{8} v^3 \, \mathcal{C}_{\phi,2}+\frac{5}{6} v^3 \, \mathcal{C}_{\phi,3},~~~~~~~~
g^{(2)}_{HHH} = v \, \mathcal{C}_{\phi,2}.
\end{eqnarray}

\vskip 5mm
\section{Description of calculations }
\label{section-III}
\par
\begin{figure*}
\begin{center}
\includegraphics[scale=0.7]{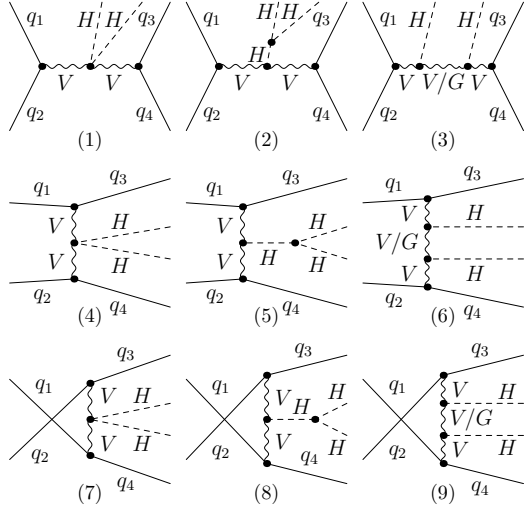}
\caption{\label{fig1} The topologies of Higgs pair production at parton level at a proton-proton collider. The diagrams with external Higgs line exchange in Figs.\ref{fig1}(3), (6) and (9) are not drawn.  }
\end{center}
\end{figure*}

Now we calculate the effects of the dimension-six operators $\mathcal{O}_{BB}$, $\mathcal{O}_{WW}$,  $\mathcal{O}_{\Phi,2}$ and $\mathcal{O}_{\Phi,3}$, which govern the VBF Higgs pair production at the LHC. The precision analysis of the VBF Higgs pair production process shows that the NLO QCD correction is of the order of $5-10\%$ and reduces the scale uncertainty of the LO integrated cross section to a few percent \cite{hpair-tot,vbfnnlo}. In our analysis the NLO QCD corrections are included for the purpose of more precise predictions. We perform the LO and NLO QCD calculation by using our developed {\sc MadGraph5} package \cite{program-2}, in which we add the codes for calculating in the EFT theory. The Feynman rules and model files of the EFT Lagrangian are obtained using FeynRules \cite{program-3}. The relevant coupling constants in the EFT are listed Eqs.(\ref{Rule-vertex-1}), (\ref{Rule-vertex-2}) and (\ref{Rule-vertex-3}). Fig.\ref{fig1} shows the topologies for pure electroweak (EW) production of $HH + 2~ {\rm jets}$ in parton level at proton-proton colliders \cite{hpair-tot}, where $V=(W,~Z, \gamma$), and $G$ denotes the Goldstone boson $G^+$ or $G^0$. In Figs.\ref{fig1}(1-3), the Higgs pair are produced in $s$-channel, known as double Higgs-strahlung mechanism. The VBF Higgs pair production mechanism is shown in Figs.\ref{fig1}(4-9), where Higgs pair can be produced through either $t$-channel or $u$-channel. The clear experimental signature of two widely separated jets and two centrally produced Higgs bosons of the VBF mechanism offers a good background suppression \cite{vbfsig}. The $s$-channel contribution tends to be sufficiently suppressed by imposing the VBF cuts \cite{vbfcut1,vbfcut2}. In order to significantly suppress the backgrounds of the VBF Higgs pair production signature, we apply the typical VBF cuts as \cite{vbfcut1}
\begin{eqnarray}\label{VBFcut}
p_{Tj} \geq 20~{\rm GeV},~~~ |y_j| \leq 4.5,~~~ y_{j_1} y_{j_2} < 0,~~~
\Delta y_{j_1j_2}=|y_{j_1}-y_{j_2}|>4,~~~ M_{j_1j_2} > 600~{\rm GeV},
\end{eqnarray}
where $p_{Tj}$ and $y_j$ are the transverse momentum and rapidity of the final jet, and $M_{j_1j_2}$ is the invariant mass of the final two jets.

\vskip 5mm
\section{Numerical results and discussion }
\label{section-IV}
\par
\subsection{Input parameters}
\par
We adopt the five-flavor scheme and neglect the masses of $u$-, $d$-, $c$-, $s$-, $b$-quark in initial parton convolution. From Eq.(\ref{mhvl}) we obtain
\begin{eqnarray}
\lambda=\frac{2}{v^2}\Big( M^2_H + M^2_H v^2 \mathcal{C}_{\Phi,2} + v^4\mathcal{C}_{\Phi,3} \Big),
\end{eqnarray}
where $v$ has the same value as in the SM. In the numerical calculation, we use the MSTW2008($68\%$ C.L.) PDFs \cite{mstw}. The SM input parameters are taken as \cite{pdg}
\begin{eqnarray}\label{SM-parameters}
&&
M_Z = 91.1876~{\rm GeV},~~~M_W = 80.385~{\rm GeV},~~~M_H = 125.09~{\rm GeV},~~~~~ \nonumber \\
&&
G_{\mu} = 1.1663787 \times 10^{-5}~{\rm GeV}^{-2},
\end{eqnarray}
and thus $v = \left(\sqrt{2} G_{\mu}\right)^{-\frac{1}{2}} = 246.22~{\rm GeV}$. In this paper, we set the factorization and renormalization scales to be equal, i.e., $\mu_F=\mu_R=\mu$. For the VBF process, we usually adopt the double-valued scale scheme in the LO calculation, i.e., set $\mu = Q_1$ and $Q_2$, the momentum transfers of the two $t$/$u$-channel weak gauge-bosons radiated off the two quark lines, in the PDF convolutions of the two incoming protons, respectively. However, this scale choice would be unambiguous at the QCD NLO due to the presence of the non-VBF virtual contribution \footnote{This non-VBF contribution is sufficiently small if the VBF cuts are applied.}. Therefore, analogous to Ref.\cite{scale_choice}, we adopt a single-valued scale defined as
\begin{eqnarray}
\label{centre-mu}
\mu(p_{T}^{H_1}, p_T^{H_2})= \left[\frac{M_H}{2} \sqrt{\frac{M_H^2}{2} + \left(p_{T}^{H_1}\right)^2 +\left(p_{T}^{H_2}\right)^2}\right]^{1/2}
\end{eqnarray}
in our calculation. This dynamical scale is dependent on the transverse momenta of the two final Higgs bosons and close to $\sqrt{Q_1Q_2}$. For comparison, we calculate the integrated cross sections and the transverse momentum distributions of the two final Higgs bosons at the QCD NLO by adopting the structure function approach \cite{sfa} \footnote{The scales $Q_1$ and $Q_2$ can be properly defined in the structure function approach at both LO and QCD NLO.} and setting $\mu=\sqrt{Q_1Q_2}$ and $\mu=\mu(p_{T}^{H_1}, p_T^{H_2})$ separately, and find that the differences from the two scale choices are less than $2\%$.

\par
Recently, an analysis of the LHC Run I measurements related to the Higgs and electroweak gauge sector combining with the triple gauge vertex results from LEP data in the framework of an effective Lagrangian has been given in Ref.\cite{constraint}. There the authors update the constraints on the coefficients of dimension-six operators, and present the allowed $95\%$ C.L. ranges of the coefficients $\mathcal{C}_{BB}$, $\mathcal{C}_{WW}$ and $\mathcal{C}_{\Phi,2}$ as
\begin{equation}\label{coefficient-range}
\mathcal{C}_{BB} \in [-3.3,\, 6.1]~{\rm TeV}^{-2},~~~\mathcal{C}_{WW} \in [-3.1,\, 3.7]~{\rm TeV}^{-2},~~~ \mathcal{C}_{\Phi,2} \in [-7.2,\, 7.5]~{\rm TeV}^{-2}.
\end{equation}
There they do not introduce ad-hoc form factors to dampen the scattering amplitude at high energies because they verified that there is no partial-wave unitarity violation in the different channels for the values of the Wilson coefficients in the $95\%$ C.L. allowed regions, except for very large and already ruled out values of $f_B$. The operator $\mathcal{O}_{\Phi,3}$ only affects the Higgs self-couplings which are largely untested so far, and therefore the coefficient $\mathcal{C}_{\Phi,3}$ is essentially unconstrained \cite{selfdect,eft-4}.
In our analysis we follow the way used in Ref.\cite{constraint} without introducing form factors.

\par
\subsection{Integrated cross sections}
\label{subsection-I}
\par
We calculate the LO integrated cross sections for the $HH + 2~{\rm jets}$ in the SM at the $14~{\rm TeV}$ LHC with the scale $\mu=M_H$, VBF cut constraints of Eq.(\ref{VBFcut}) and the input parameters in Eq.(\ref{SM-parameters}). The LO cross section contributed by all the $s$-, $t$- and $u$-channel diagrams in Fig.\ref{fig1} is obtained as $\sigma^{(s+t+u)}= 0.808(2)~fb$, while the cross section from only the $t$- and $u$-channel diagrams (i.e., the VBF diagrams) is $\sigma_{\rm cuts}^{(t+u)}= 0.807(2)~fb$. We see that those two numerical results are in good agreement with each other within the calculation errors. It ensures that we can ignore the $s$-channel diagrams, i.e., consider only the VBF diagrams, if the VBF cuts of (\ref{VBFcut}) are applied.

\par
In order to check the correctness of our calculations, we perform the QCD NLO calculations of the VBF Higgs pair production process at the $14~{\rm TeV}$ LHC in the SM by taking the input parameters as listed in Eq.(\ref{SM-parameters}), the constraints in Eq.(\ref{VBFcut}) and $\mu=M_H$. We obtain our numerical results as $\sigma_{NLO}= 0.867(2)~fb$ by using {\sc MadGraph5} package \cite{program-2}, and $\sigma_{NLO}= 0.864(2)~fb$ by applying {\sc VBFNLO} program \cite{VBFNLO}. Both numerical results are consistent with each other within statistic errors.

\par
To estimate missing scale uncertainty of the integrated cross section for the VBF Higgs pair production process, we set $\mu=\kappa \mu_0$ where $\mu_0=\mu(p_{T}^{H_1}, p_T^{H_2})$, and vary $\kappa$ in the range of $[1/4,~4]$ as suggested in Ref.\cite{vbfnnlo}. By taking the VBF cuts shown in Eq.(\ref{VBFcut}) we obtain the central values of the LO and NLO corrected integrated cross sections at $\kappa=1$ with the errors due to scale uncertainty at the $14~TeV$ LHC as $\sigma_{LO}^{SM}=0.822^{+0.166}_{-0.123}~fb$, and $\sigma_{NLO}^{SM}=0.878^{+0.010}_{-0.019}~fb$ separately. We see that the scale uncertainty of NLO is much smaller than the LO. In following analysis we shall focus on the discussion based on the QCD NLO corrected results by taking $\mu=\mu_0$ unless stated otherwise.

\par
As mentioned in section \ref{section-II}, only the dimension-six operators in the following reduced sets are considered for the VBF Higgs pair production at the LHC:
\begin{eqnarray}
\label{reduced-set}
\overline{B} = \{ \mathcal{O}_{BB},~ \mathcal{O}_{WW} \},~~~~~~~~~~\overline{C} = \{ \mathcal{O}_{\Phi,2},~ \mathcal{O}_{\Phi,3} \}\,.
\end{eqnarray}
The constraints on the coefficients $\mathcal{C}_{BB}$ and $\mathcal{C}_{WW}$ are substantially more stringent than on $\mathcal{C}_{\Phi,2}$ and $\mathcal{C}_{\Phi,3}$ \cite{constraint}, and the latter two coefficients directly modify the Higgs self-couplings. In this subsection, we investigate the influence on the integrated cross section from each of the four dimension-six operators, and also analyze the joint effect from the two operators in each reduced set in Eq.(\ref{reduced-set}).

\par
\begin{figure*}
\begin{center}
\includegraphics[scale=0.7]{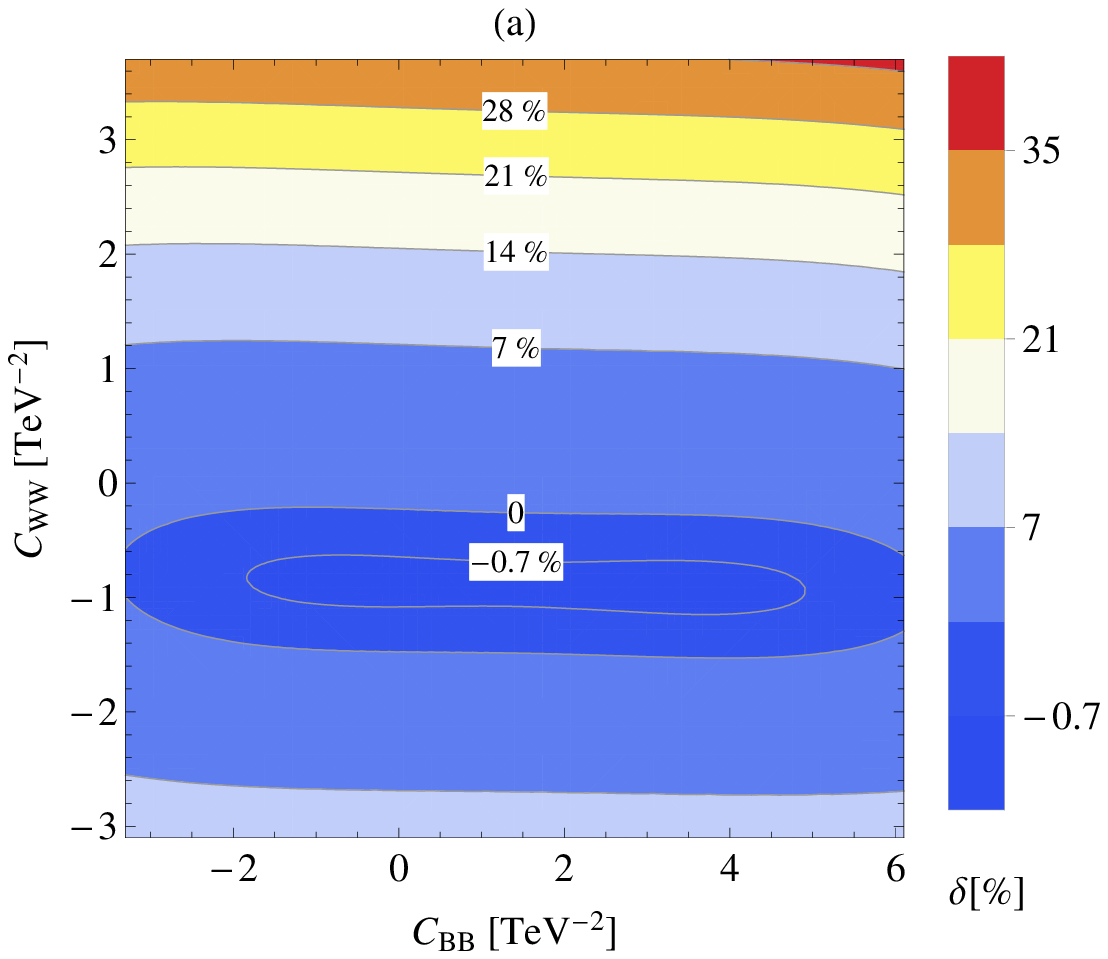}
\includegraphics[scale=0.7]{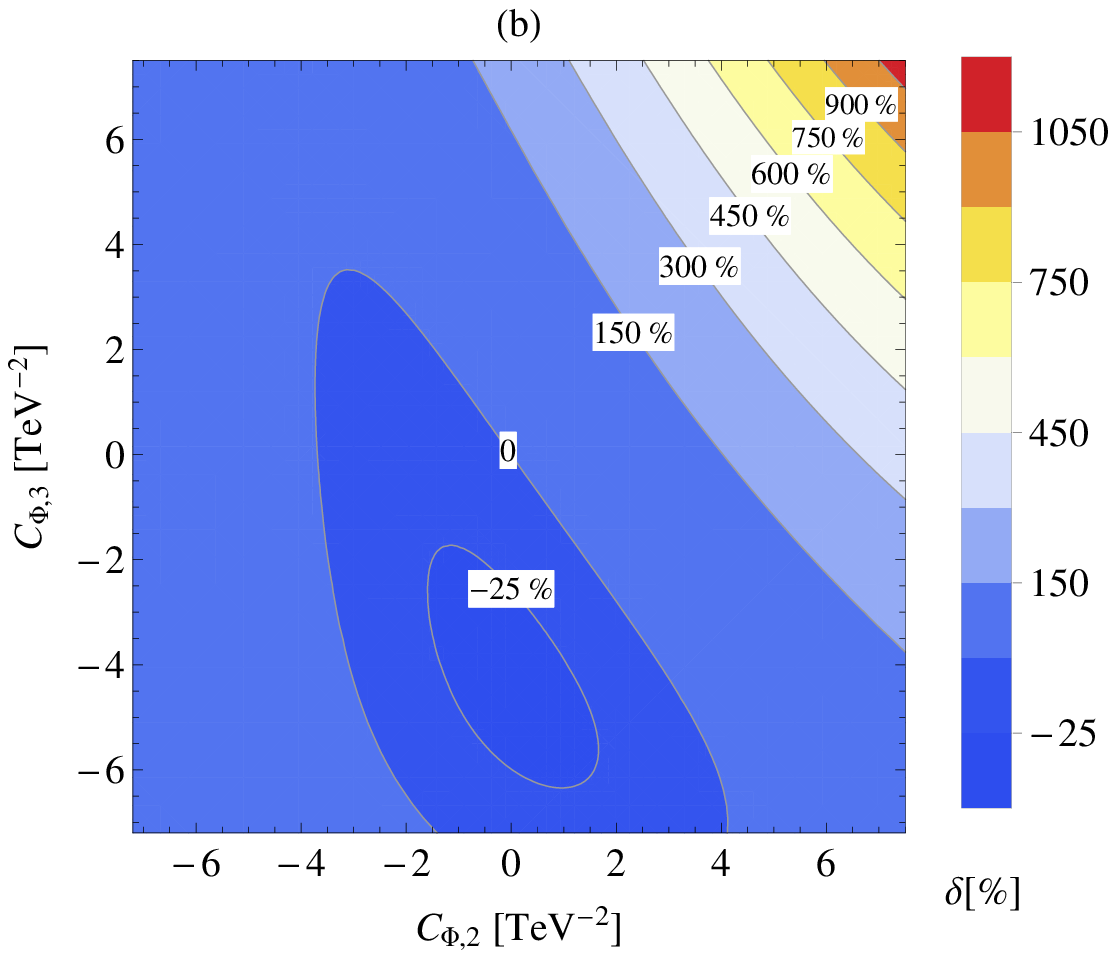}
\caption{\label{fig2} The dependence of the relative discrepancy $\delta({\mathcal{C}_i,\mathcal{C}_j})$ on the coefficients $\mathcal{C}_i$ and $\mathcal{C}_j$ for the VBF Higgs pair production at the $14~{\rm TeV}$ LHC. (a) $\mathcal{C}_i,~ \mathcal{C}_j = \mathcal{C}_{BB},~\mathcal{C}_{WW}$. (b) $\mathcal{C}_i,~ \mathcal{C}_j = \mathcal{C}_{\Phi,2},~\mathcal{C}_{\Phi,3}$.}
\end{center}
\end{figure*}

In Figs.\ref{fig2}(a) and (b) we display the contour maps of the relative discrepancies of the integrated cross section, defined as $\delta({\mathcal{C}_i,\mathcal{C}_j}) = \left[\sigma({\mathcal{C}_i,\mathcal{C}_j}) - \sigma^{\rm SM} \right]/\sigma^{\rm SM}$ (in percent), on the $\mathcal{C}_{BB}$-$\mathcal{C}_{WW}$ and $\mathcal{C}_{\Phi,2}$-$\mathcal{C}_{\Phi,3}$ planes, separately. Fig.\ref{fig2}(a) shows that most contour-lines of the relative discrepancy $\delta({\mathcal{C}_{BB},\mathcal{C}_{WW}})$ are nearly parallel to the $\mathcal{C}_{BB}$-axis, and $\delta({\mathcal{C}_{BB},\mathcal{C}_{WW}})$ can exceed $28\%$ in the whole plotted $\mathcal{C}_{BB}$ region ($-3.3~{\rm TeV^{-2}} \leqslant \mathcal{C}_{BB} \leqslant 6.1~{\rm TeV^{-2}}$) if $\mathcal{C}_{WW} \geqslant 3.4~ {\rm TeV^{-2}}$. Fig.\ref{fig2}(b) indicates that $\delta({\mathcal{C}_{\Phi,2},\mathcal{C}_{\Phi,3}})$ runs up to the value beyond $1050\%$ in the region of $(\mathcal{C}_{\Phi,2}>7.0~{\rm TeV^{-2}},~\mathcal{C}_{\Phi,3}>7.0~{\rm TeV^{-2}})$. We can see that $\delta({\mathcal{C}_{BB},\mathcal{C}_{WW}})$ is less sensitive to both $\mathcal{C}_{BB}$ and $\mathcal{C}_{WW}$, while $\delta({\mathcal{C}_{\Phi,2},\mathcal{C}_{\Phi,3}})$ is particularly sensitive to both $\mathcal{C}_{\Phi,2}$ and $\mathcal{C}_{\Phi,3}$. The variation range of $\delta({\mathcal{C}_{\Phi,2},\mathcal{C}_{\Phi,3}})$ is about 30 times larger than that of $\delta({\mathcal{C}_{BB},\mathcal{C}_{WW}})$. That means the measurement of the cross section for the VBF Higgs pair production is more suitable for ascertaining the limitation ranges of the coefficients $\mathcal{C}_{\Phi,2}$ and $\mathcal{C}_{\Phi,3}$. Figs.\ref{fig2}(a) and (b) also show that there exist some regions on the $\mathcal{C}_{BB}$-$\mathcal{C}_{WW}$ and $\mathcal{C}_{\Phi,2}$-$\mathcal{C}_{\Phi,3}$ planes, respectively, in which $\delta_{\mathcal{C}_{BB},\mathcal{C}_{WW}}$ and $\delta_{\mathcal{C}_{\Phi,2},\mathcal{C}_{\Phi,3}}$ become negative, which is due to the destructive interference between the SM and the anomalous coupling amplitudes.

\par
\begin{figure*}
\begin{center}
\includegraphics[scale=0.29]{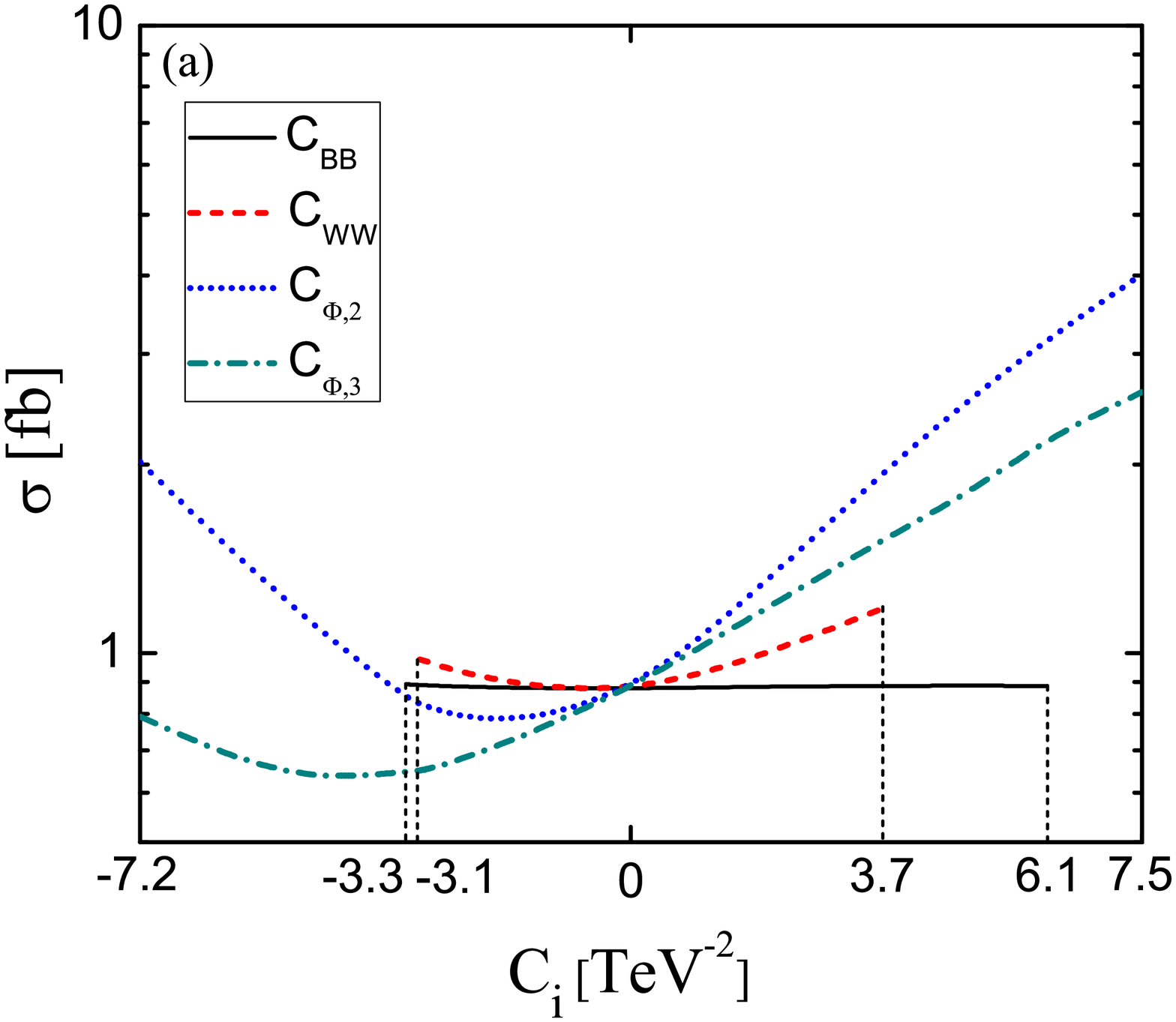}
\includegraphics[scale=0.29]{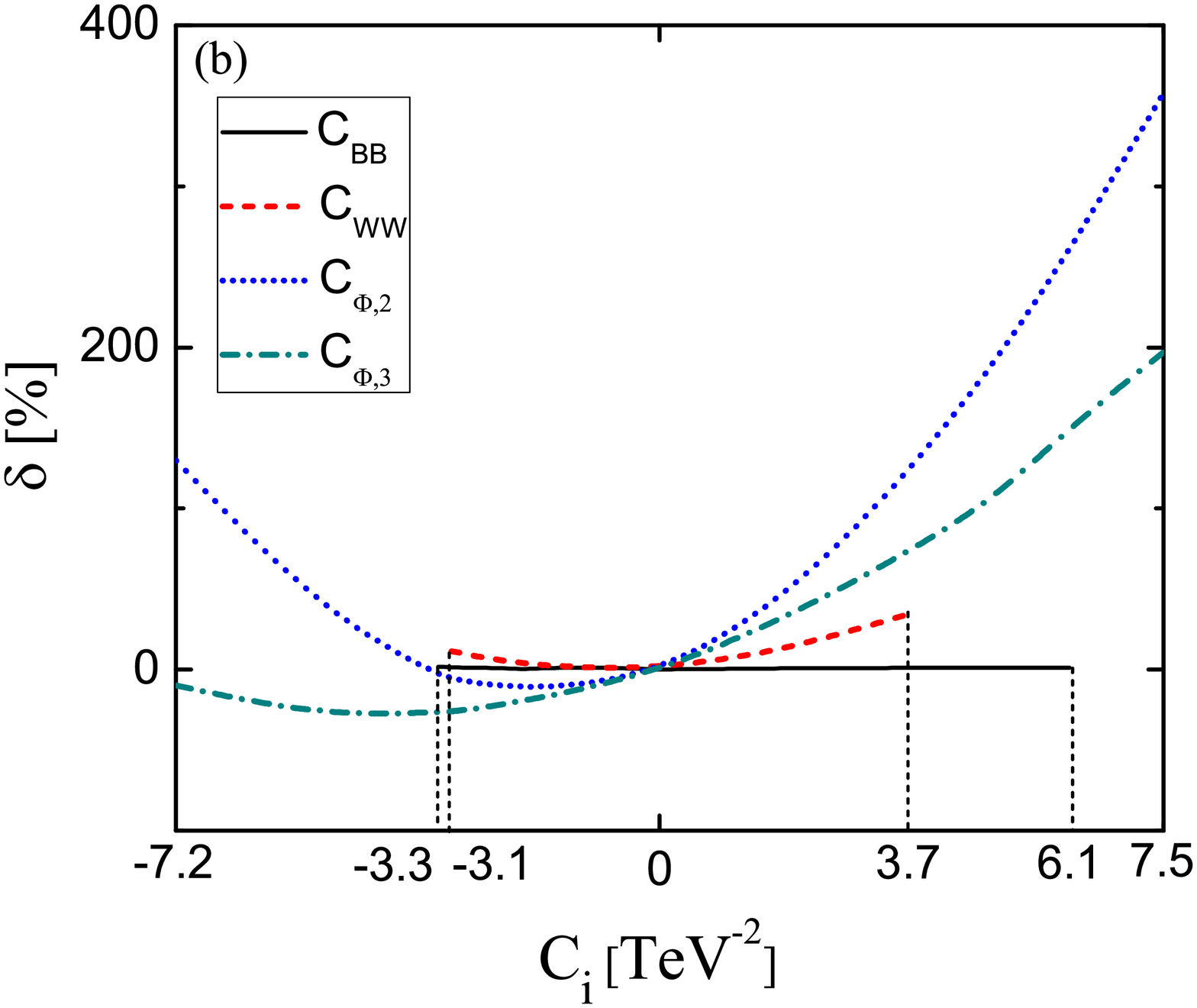}
\caption{\label{fig3}
The dependence of (a) the integrated cross section and (b) the relative discrepancy of the VBF Higgs pair production at the $14~{\rm TeV}$ LHC on $\mathcal{C}_{BB}$, $\mathcal{C}_{WW}$, $\mathcal{C}_{\Phi,2}$ and $\mathcal{C}_{\Phi,3}$ within the single operator dominance hypothesis.}
\end{center}
\end{figure*}

The dependence of the integrated cross section and the relative discrepancy $\delta_{\mathcal{C}_i}$, defined as $\delta_{\mathcal{C}_i} = ( \sigma_{\mathcal{C}_i}-\sigma^{\rm SM} )/\sigma^{\rm SM}$, on each coefficient $\mathcal{C}_{i} \in \{ \mathcal{C}_{BB},~ \mathcal{C}_{WW},~ \mathcal{C}_{\Phi,2},~ \mathcal{C}_{\Phi,3} \}$ at the $14~ {\rm TeV}$ LHC are provided in Figs.\ref{fig3}(a) and (b), respectively, where the single operator dominance hypothesis is employed, i.e., the other three coefficients $\mathcal{C}_{j \neq i}$ are set to zero in discussing the $\mathcal{C}_{i}$ dependence. The two figures clearly show that the effect from the operator $\mathcal{O}_{BB}$ is negligible and $\delta_{\mathcal{C}_{BB}} \sim 0$ in the range of $-3.3~{\rm TeV^{-2}} \leqslant \mathcal{C}_{BB} \leqslant 6.1~{\rm TeV^{-2}}$. The effect from $\mathcal{O}_{WW}$ is also heavily constrained by experiments and $|\delta_{\mathcal{C}_{WW}}| < 34.4\%$ as $\mathcal{C}_{WW} \in [-3.1,~ 3.7]~{\rm TeV}^{-2}$. However, the integrated cross section is particularly sensitive to both $\mathcal{C}_{\Phi,2}$ and $\mathcal{C}_{\Phi,3}$ separately. The relative discrepancies $\delta_{\mathcal{C}_{\Phi,2}}$ and $\delta_{\mathcal{C}_{\Phi,3}}$ increase remarkably as the increment of $\mathcal{C}_{\Phi,2}$ and $\mathcal{C}_{\Phi,3}$ when $\mathcal{C}_{\Phi,2},~ \mathcal{C}_{\Phi,3} > 0$, and reach $357.9\%$ and $197.3\%$, respectively, at $\mathcal{C}_{\Phi,2}=\mathcal{C}_{\Phi,3}=7.5~{\rm TeV}^{-2}$.

\par
To demonstrate the effects from the dimension-six operators $\mathcal{O}_{BB}$, $\mathcal{O}_{WW}$, $\mathcal{O}_{\Phi,2}$ and $\mathcal{O}_{\Phi,3}$ on the VBF Higgs pair production more clearly, we also list the numerical results of the integrated cross sections and the corresponding relative discrepancies for some typical values of $\mathcal{C}_{BB}$, $\mathcal{C}_{WW}$, $\mathcal{C}_{\Phi,2}$ and $\mathcal{C}_{\Phi,3}$ at the $\sqrt{S}=13,~14~ {\rm TeV}$ LHC within the single operator dominance hypothesis in Table \ref{tab1}. As shown in the table and above figures, we can see that the relative discrepancies caused by the operators $\mathcal{O}_{BB}$ and $\mathcal{O}_{WW}$ are generally much smaller than those by the operators $\mathcal{O}_{\Phi,2}$ and $\mathcal{O}_{\Phi,3}$ in most of the experimentally allowed regions of the corresponding coefficients $\mathcal{C}_i$. We conclude that the operators $\mathcal{O}_{BB}$ and $\mathcal{O}_{WW}$ have less influence on the VBF Higgs pair production at the LHC. Therefore, in the further numerical calculations we set $\mathcal{C}_{BB}=\mathcal{C}_{WW}=0$, and consider only the effects induced by the operators $\mathcal{O}_{\Phi,2}$ and $\mathcal{O}_{\Phi,3}$ on the integrated cross section and kinematic distributions.
\begin{table}
\newcommand{\tat}[2]{\begin{tabular}{@{}#1@{}}#2\end{tabular}}
\begin{center}
\begin{tabular}{|cccc|cc|cc|}
\hline \multirow{2}{*}{ \tat{c}{$\mathcal{C}_{BB}$\\$[{\rm TeV^{-2}}]$} }
      &\multirow{2}{*}{ \tat{c}{$\mathcal{C}_{WW}$\\$[{\rm TeV^{-2}}]$} }
      &\multirow{2}{*}{ \tat{c}{$\mathcal{C}_{\Phi,2}$\\$[{\rm TeV^{-2}}]$} }
      &\multirow{2}{*}{ \tat{c}{$\mathcal{C}_{\Phi,3}$\\$[{\rm TeV^{-2}}]$} }
      & \multicolumn{2}{c|}{$\sqrt{S}=13~{\rm TeV}$}  & \multicolumn{2}{c|}{$\sqrt{S}=14~{\rm TeV}$} \\
      \cline{5-8} & & & &
      $\sigma_{NLO}~[fb]$ & $\delta~[\%]$ & $\sigma_{NLO}~[fb]$ & $\delta~[\%]$                    \\
\hline
$-3.3~~\,$  & 0          & 0   & 0   & 0.745(3)  &  0.5    & 0.892(3) & 1.6    \\
$-1.0~~\,$  & 0          & 0   & 0   & 0.739(3)  &  $-0.3$ & 0.884(3) & 0.7    \\
$6.1$       & 0          & 0   & 0   & 0.751(3)  &  1.3    & 0.887(3) & 1.0    \\
\hline  0   &$-3.1~~\,$  & 0   & 0   & 0.811(3)  &  9.4    & 0.978(4) & 11.4   \\
        0   &$-1.0~~\,$  & 0   & 0   & 0.735(3)  &  $-0.8$ & 0.875(3) & $-0.3$ \\
        0   & $3.7$      & 0   & 0   & 0.964(4)  &  30.1   & 1.180(4) & 34.4   \\
\hline  0   & 0   & $-7.2~~\,$ & 0   & 1.618(7)  &  118.4  & 2.018(9) & 129.8  \\
        0   & 0   & $-2.0~~\,$ & 0   & 0.623(3)  &  $-15.9$& 0.763(3) & $-13.1$\\
        0   & 0   & $7.5\,$    & 0   & 3.34(1)   &  350.7  & 4.02(2)  & 357.9  \\
\hline  0   & 0   & 0   & $-7.2~~\,$ & 0.660(3)  &  $-10.9$& 0.792(3) & $-9.8$ \\
        0   & 0   & 0   & $-4.0~~\,$ & 0.524(2)  &  $-29.3$& 0.637(3) & $-27.4$\\
        0   & 0   & 0   &  $7.5\,$   & 2.21(1)   &  198.2  & 2.61(1)  & 197.3  \\
\hline  0   & 0   & 0   &  0         & 0.741(3)  &  0      & 0.878(3) & 0      \\

\hline
\end{tabular}
\caption{
\label{tab1}
The NLO QCD corrected integrated cross sections and the corresponding relative discrepancies of the VBF
Higgs pair production at the $13$ and $14~{\rm TeV}$ LHC for some typical values of $\mathcal{C}_{BB}$,
$\mathcal{C}_{WW}$, $\mathcal{C}_{\Phi,2}$ and $\mathcal{C}_{\Phi,3}$ within the single operator dominance hypothesis.}
\end{center}
\end{table}

\par
\begin{figure*}
\begin{center}
\includegraphics[scale=0.29]{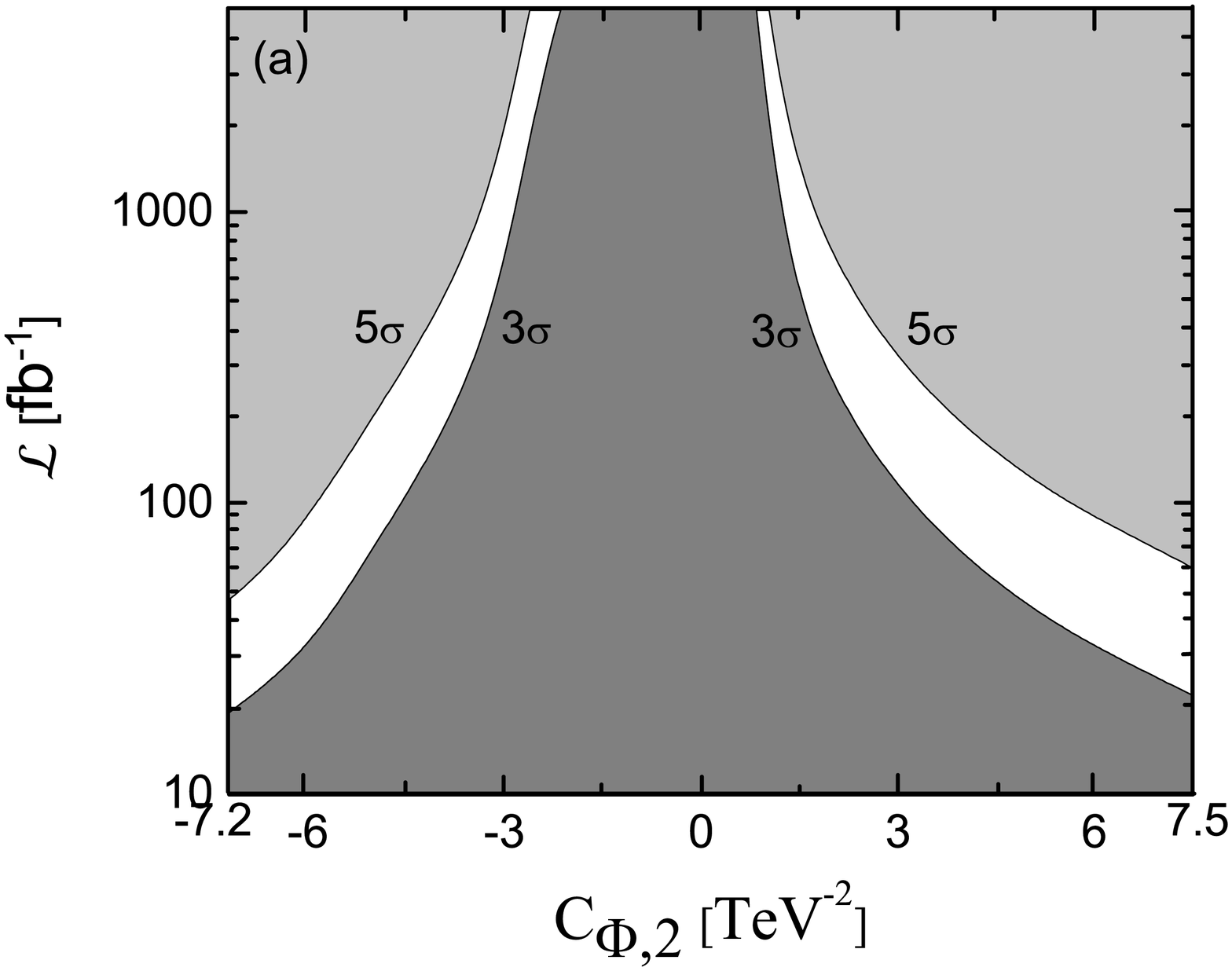}
\includegraphics[scale=0.29]{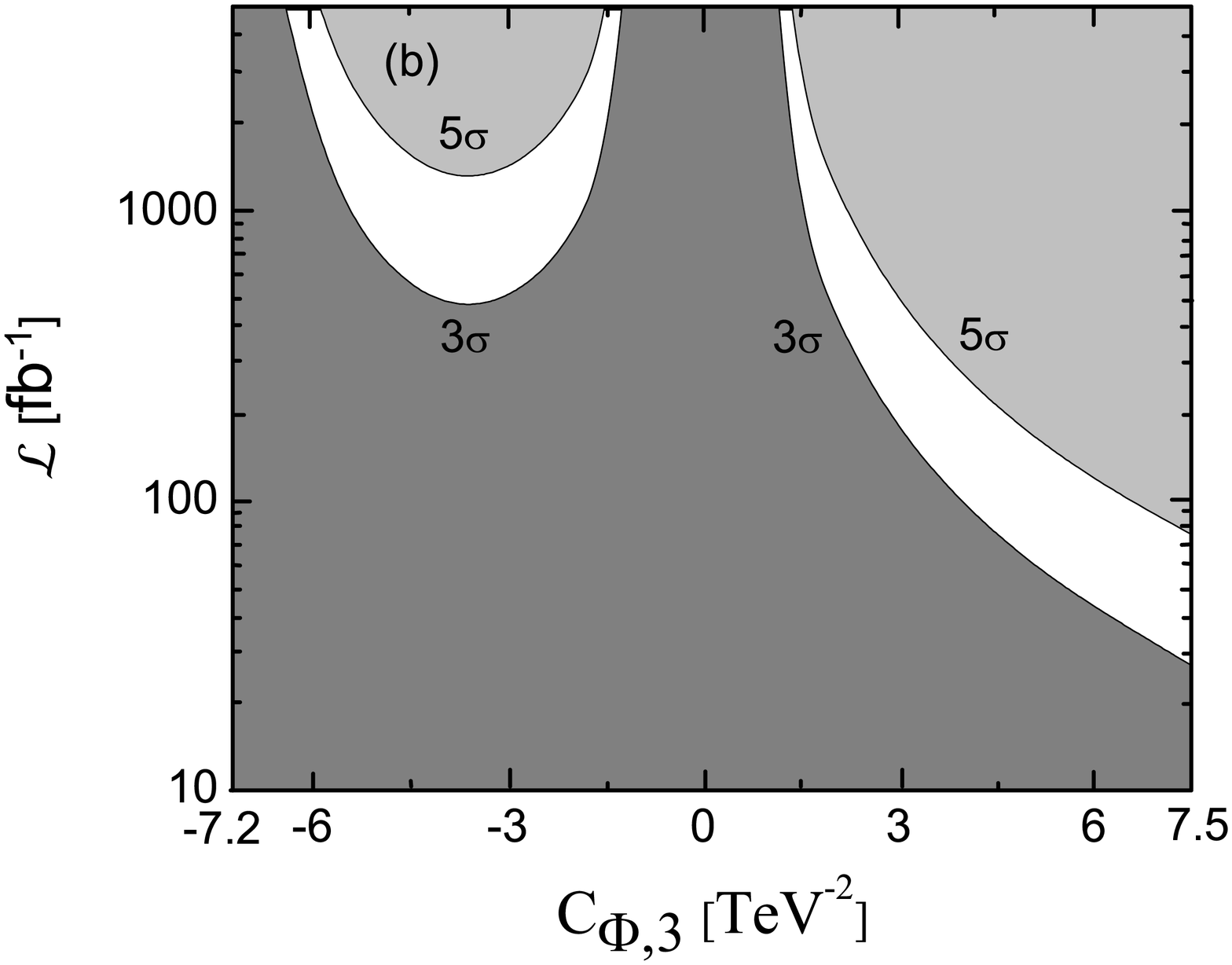}
\caption{
\label{fig4}
The $5\sigma$ observation area (light gray) and the $3\sigma$ exclusion area (dark gray) for the VBF process $pp \to HH+X \to bb\bar{b}\bar{b}+X$ with the cuts of Eqs.($\ref{decaybcut-1}$) and ($\ref{decaybcut-2}$) at $\sqrt{S}=14$ LHC. (a) in the $\mathcal{L}$-$\mathcal{C}_{\Phi,2}$ space. (b) in the $\mathcal{L}$-$\mathcal{C}_{\Phi,3}$ space. }
\end{center}
\end{figure*}

As mentioned in the above discussion, the $\mathcal{O}_{\Phi,2}$ and $\mathcal{O}_{\Phi,3}$ dimension-six operators, which are relevant to the VBF Higgs pair production process at the LHC, can obviously modify the SM predicted integrated cross section, if the new physics really exists. To further explore the discovery and exclusion potential for the signals of the two operators, we adopt the 5$\sigma$ discovery and 3$\sigma$ exclusion limits to study the constraints on the coefficients $\mathcal{C}_{\Phi,2}$ and $\mathcal{C}_{\Phi,3}$. We assume that the new physics effect can be discovered and excluded if
\begin{eqnarray}\label{upper}
\Delta \sigma(\mathcal{C}_{\Phi,i}) \geq 5\times\frac{\sqrt{\sigma_{NLO}(\mathcal{C}_{\Phi,i})\mathcal{L}}}{\mathcal{L}}
\end{eqnarray}
and
\begin{eqnarray}\label{lower}
\Delta \sigma(\mathcal{C}_{\Phi,i}) \leq 3\times\frac{\sqrt{\sigma_{NLO}(\mathcal{C}_{\Phi,i})\mathcal{L}}}{\mathcal{L}},
\end{eqnarray}
respectively, where $\Delta \sigma(\mathcal{C}_{\Phi,i})=\left|\sigma_{NLO}(\mathcal{C}_{\Phi,i}) - \sigma_{NLO}^{SM}\right|,~ (i=2,3)$ and $\mathcal{L}$ is the integrated luminosity. We discuss the $5\sigma$ discovery and $3\sigma$ exclusion limits via measuring the cross section of the VBF Higgs pair production process including the sequential on-shell Higgs-boson decays of $H \to b\bar{b}$. We define the leading b-jet ($b_1$-jet) as the one with the largest transverse momentum among the final four b-jets, In the discussion we collect the events by taking the constraint for the leading b-jet as
\begin{eqnarray}\label{decaybcut-1}
200~{\rm GeV} \leq p_{T,b_1} \leq 600~{\rm GeV},~~~ |y_{b_1}| \leq 4.5~.
\end{eqnarray}
and the transverse momentum and rapidity limits for the other three b-jets as
\begin{eqnarray}\label{decaybcut-2}
p_{T,b} \geq 20~{\rm GeV},~~~ |y_b| \leq 4.5~.
\end{eqnarray}
Assuming the branch ratio of the Higgs boson decay $H \to b \bar{b}$ having the value same as in the SM, and we obtain $Br(H \to b \bar{b})= 60.70\%$ by adopting the program HDECAY \cite{hdecay} with the input  parameters from Eq.(\ref{SM-parameters}). Figs.\ref{fig4} (a) and (b) show the 5$\sigma$ discovery and 3$\sigma$ exclusion regions on the $\mathcal{L}$-$\mathcal{C}_{\Phi,2}$ and $\mathcal{L}$-$\mathcal{C}_{\Phi,3}$ planes for the VBF Higgs pair production including $H \to b \bar{b}$ decays with the cut conditions of ($\ref{decaybcut-1}-\ref{decaybcut-2}$) separately. In both figures the integrated luminosity $\mathcal{L}$ is plotted in the ranges from $10 fb^{-1}$ to $5000 fb^{-1}$, and the light gray and dark gray regions represent the parameter space where the effect of the corresponding dimension-six operator can and cannot be observed, separately. We can read out from  Fig.\ref{fig4}(a) that, the $\mathcal{O}_{\Phi,2}$ effect can be observed when $\mathcal{C}_{\Phi,2} < -3.8~{\rm TeV^{-2}}$ or $\mathcal{C}_{\Phi,2}>2.2~{\rm TeV^{-2}}$, and cannot be observed when $-3.0~{\rm TeV^{-2}} < \mathcal{C}_{\Phi,2} < 1.5~{\rm TeV^{-2}}$, for $\mathcal{L}=600 fb^{-1}$. We can also read out from Fig.\ref{fig4}(b) that for $\mathcal{L}=600 fb^{-1}$, the $\mathcal{O}_{\Phi,3}$ effect can be observed when $\mathcal{C}_{\Phi,3} >2.8~{\rm TeV^{-2}}$, and cannot be observed when $\mathcal{C}_{\Phi,3} < -4.7~{\rm TeV^{-2}}$ or $-2.7~{\rm TeV^{-2}} < \mathcal{C}_{\Phi,3} < 1.7~{\rm TeV^{-2}}$.

\par
\subsection{Kinematic distributions}\label{subsection-II}
\par
Now we study the influences of the dimension-six operators $\mathcal{O}_{\Phi,2}$ and $\mathcal{O}_{\Phi,3}$ on the kinematic distributions of the VBF Higgs pair production at the $14~{\rm TeV}$ LHC within the single operator dominance hypothesis. The relative discrepancy of the normalized differential cross section with respect to the kinematic variable $X$ is defined as
\begin{eqnarray}
\delta_{\Phi,i}(X)\equiv\left(\frac{1}{\sigma_{\Phi,i}} \frac{d\sigma_{\Phi,i}}{dX}
                          -\frac{1}{\sigma^{\rm SM}} \frac{d\sigma^{\rm SM}}{dX} \right)
              \left/\left(\frac{1}{\sigma^{\rm SM}} \frac{d\sigma^{\rm SM}}{dX}  \right),~~~~~~~(i=2,3)\,.\right.
\end{eqnarray}

\par
\begin{figure*}
\begin{center}
\includegraphics[scale=0.29]{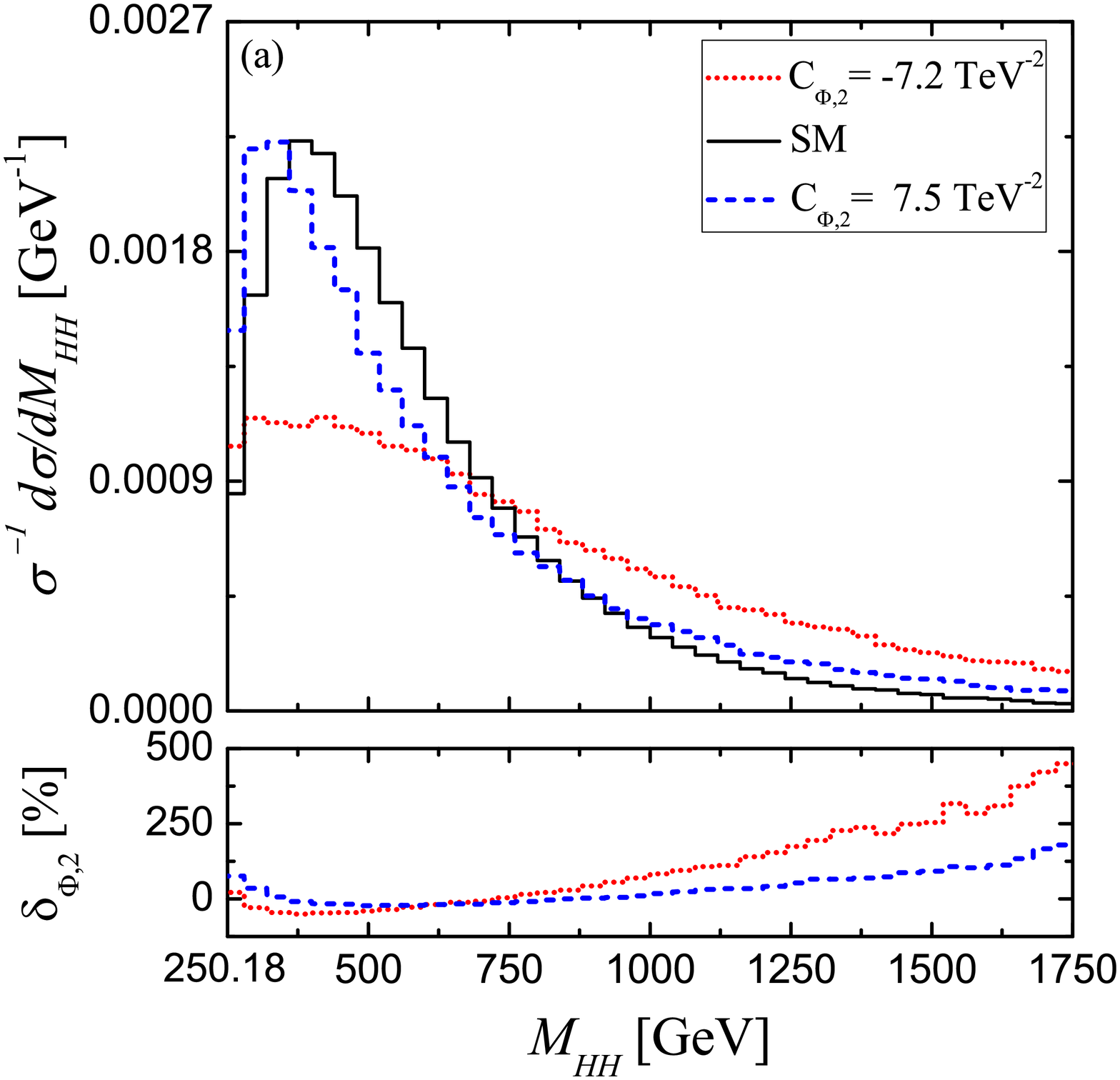}
\includegraphics[scale=0.29]{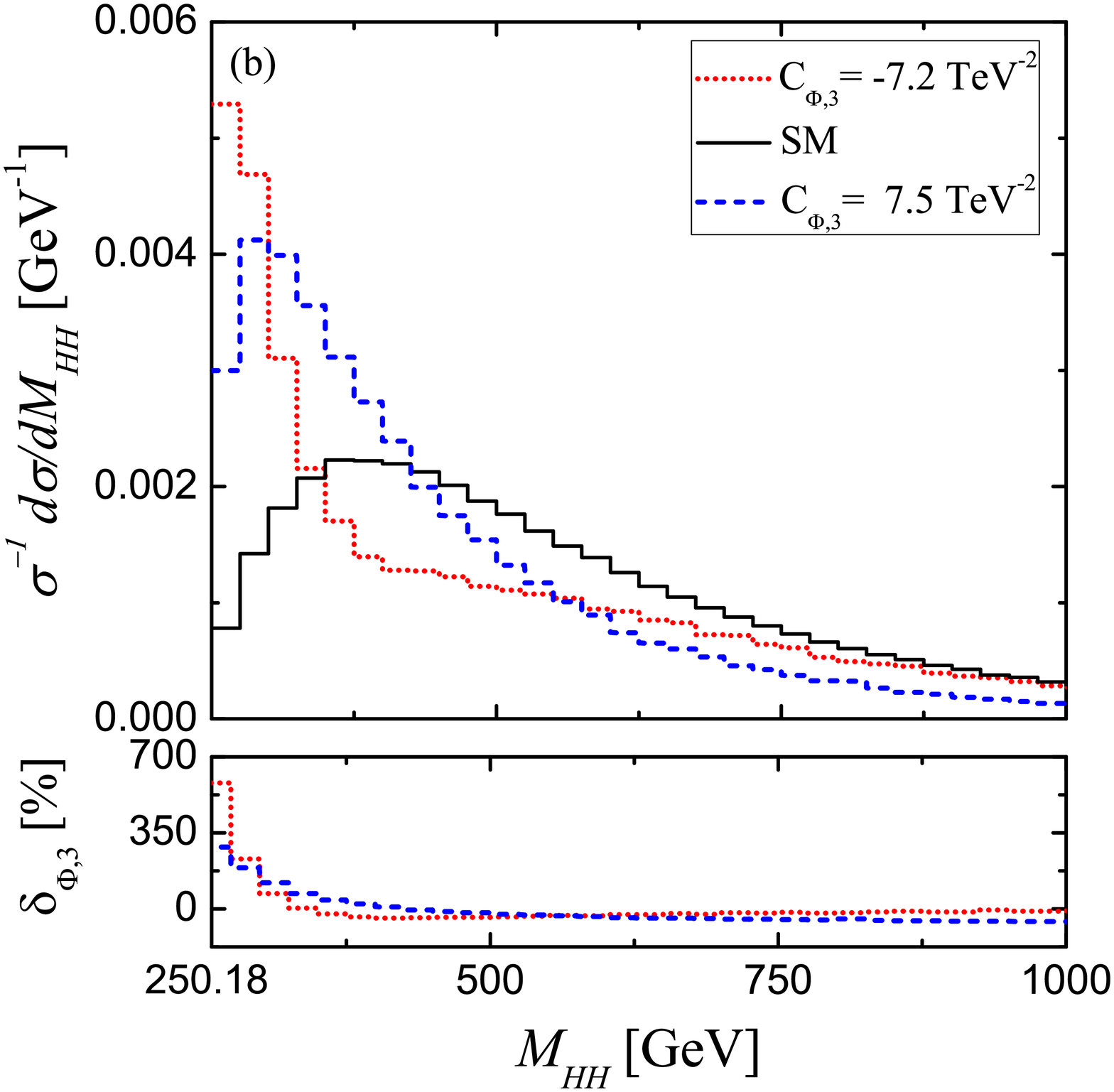}
\caption{
\label{fig5}
The normalized distributions of the Higgs pair invariant mass and the corresponding relative discrepancies for the VBF Higgs pair production at the $14~ {\rm TeV}$ within the single operator dominance hypothesis. (a) the three curves are for the SM (black full line), $\mathcal{C}_{BB} = \mathcal{C}_{WW} = \mathcal{C}_{\Phi,3} = 0$, $\mathcal{C}_{\Phi,2}=-7.2~{\rm TeV^{-2}}$ (red dotted line) and $\mathcal{C}_{BB} = \mathcal{C}_{WW} = \mathcal{C}_{\Phi,3} = 0$, $\mathcal{C}_{\Phi,2}=7.5~{\rm TeV^{-2}}$ (blue dashed line). (b) the three curves are for the SM (black full line), $\mathcal{C}_{BB} = \mathcal{C}_{WW} = \mathcal{C}_{\Phi,2} = 0$, $\mathcal{C}_{\Phi,3}=-7.2~{\rm TeV^{-2}}$ (red dotted line) and $\mathcal{C}_{BB} = \mathcal{C}_{WW} = \mathcal{C}_{\Phi,2} = 0$, $\mathcal{C}_{\Phi,3}=7.5~{\rm TeV^{-2}}$ (blue dashed line). }
\end{center}
\end{figure*}

In Figs.\ref{fig5}(a) and (b) we depict the normalized distributions of the Higgs pair invariant mass, i.e., $\frac{1}{\sigma}\frac{d \sigma}{d M_{HH}}$, at the $14~{\rm TeV}$ LHC for the operators $\mathcal{O}_{\Phi,2}$ and $\mathcal{O}_{\Phi,3}$, separately. The relative discrepancies $\delta_{\Phi,2}(M_{HH})$ and $\delta_{\Phi,3}(M_{HH})$ are depicted in the nether plots of Figs.\ref{fig5} (a) and (b), respectively. We see from these figures that the two operators, $\mathcal{O}_{\Phi,2}$ and $\mathcal{O}_{\Phi,3}$, substantially change the shapes of the $M_{HH}$ distributions correspondingly. The nether plot of Fig.\ref{fig5}(a) shows that $\delta_{\Phi,2}(M_{HH})$ for $\mathcal{C}_{\Phi,2}=7.5$ and $-7.2~{\rm TeV^{-2}}$ decrease gradually until they reach their minima at $M_{HH} \sim 550$ and $400~{\rm GeV}$, respectively, and then increase further with the increment of $M_{HH}$. In contrast, $\delta_{\Phi,3}(M_{HH})$ for both $\mathcal{C}_{\Phi,3} = 7.5 ~{\rm TeV^{-2}}$ and $\mathcal{C}_{\Phi,3} = -7.2~{\rm TeV^{-2}}$ shown in the nether panel of Fig.\ref{fig5}(b) reach their maxima at $M_{HH} = 2 M_H$. We may conclude that the Higgs pair has more possibilities to be produced in high (low) $M_{HH}$ region via the VBF at the $14~ {\rm TeV}$ LHC compared to the SM prediction if the dimension-six operator $\mathcal{O}_{\Phi,2}$ ($\mathcal{O}_{\Phi,3}$) dominates the new physics beyond the SM. The shape of the $M_{HH}$ distribution for the VBF Higgs pair production is helpful for determining the source of the new physics in the EFT.

\par
\begin{figure*}
\begin{center}
\includegraphics[scale=0.29]{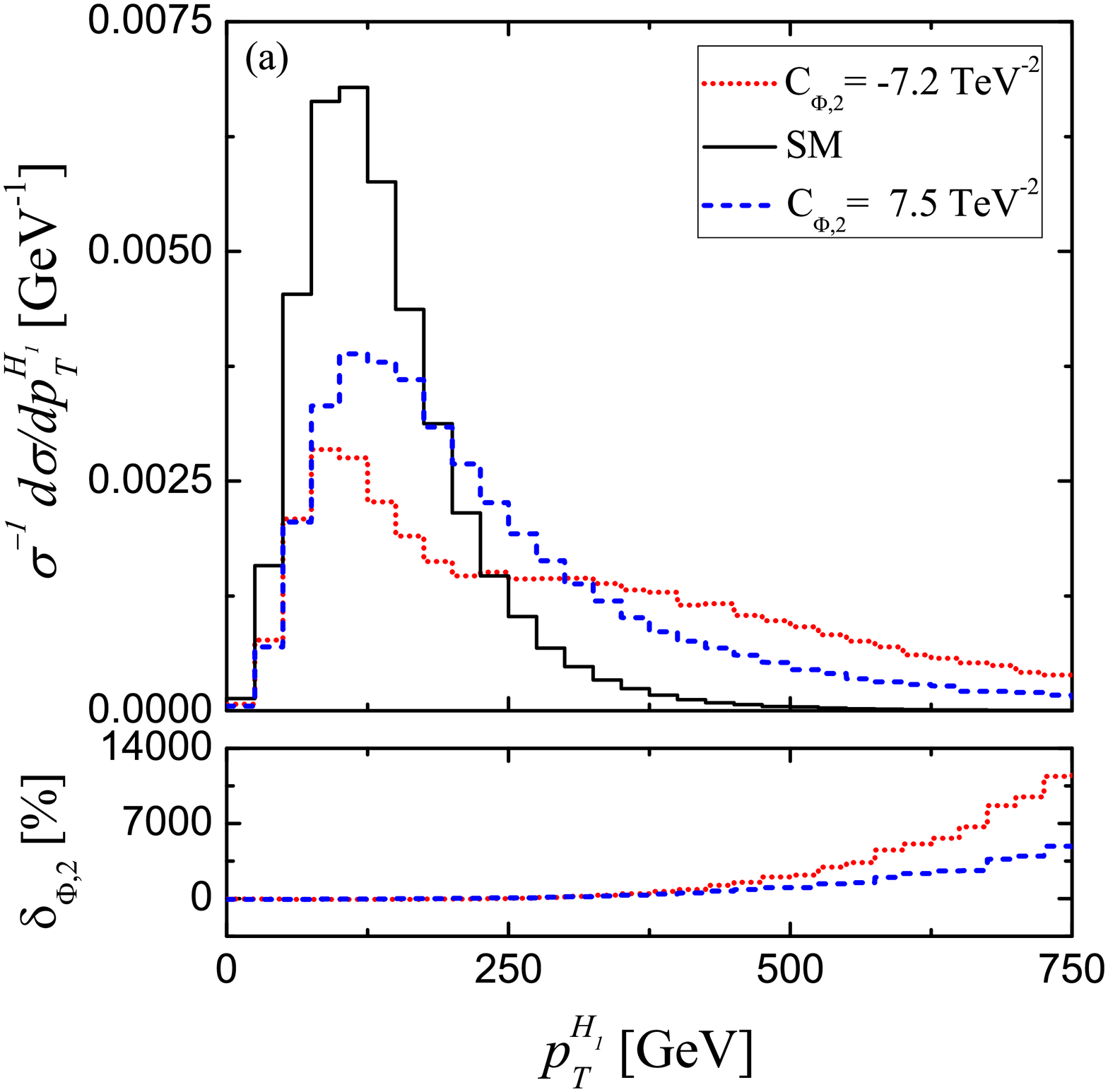}
\includegraphics[scale=0.29]{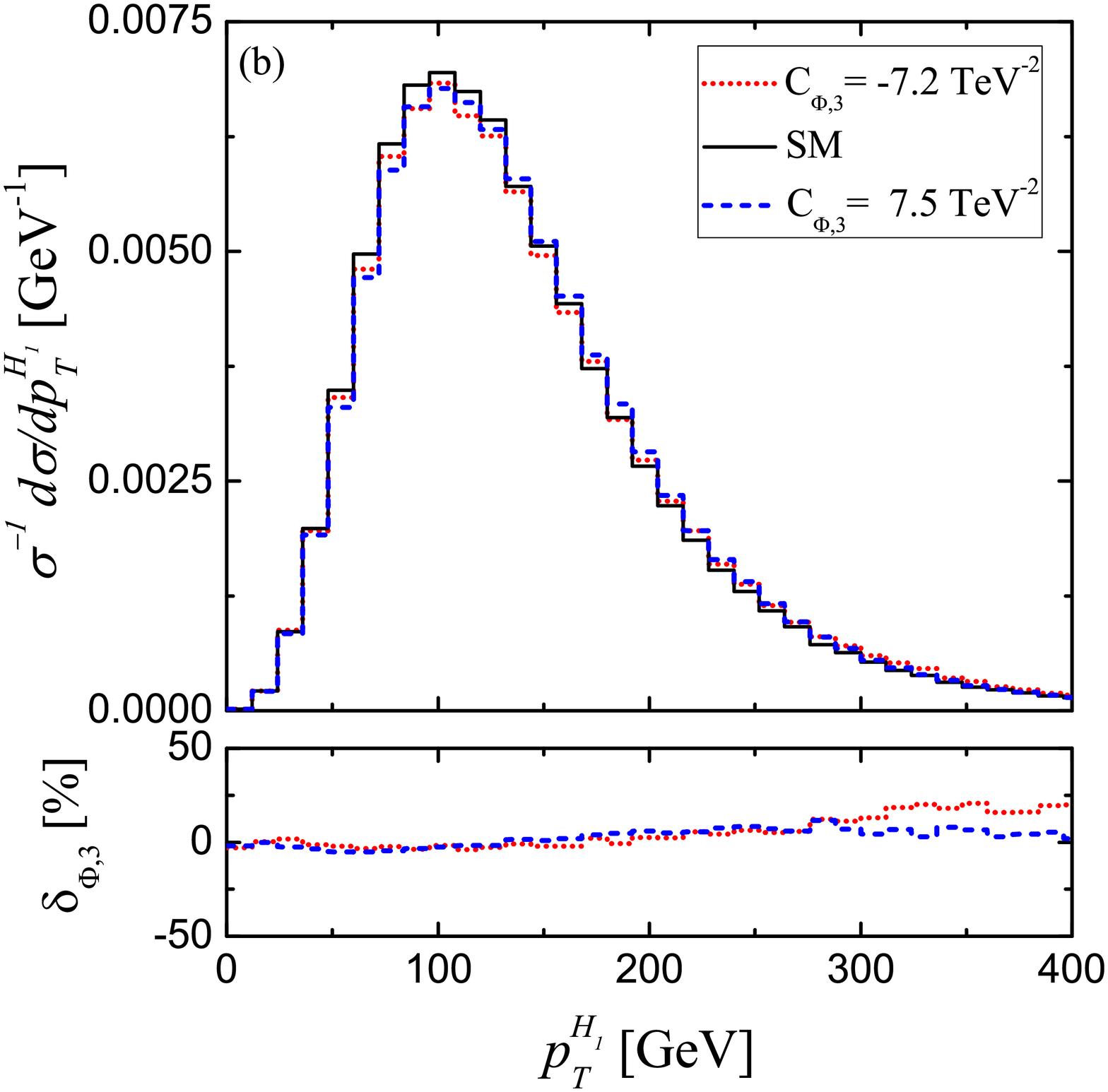}
\caption{
\label{fig6}
The same as Fig.\ref{fig5} but for the transverse momentum distributions of $H_1$. }
\end{center}
\end{figure*}
\begin{figure*}
\begin{center}
\includegraphics[scale=0.29]{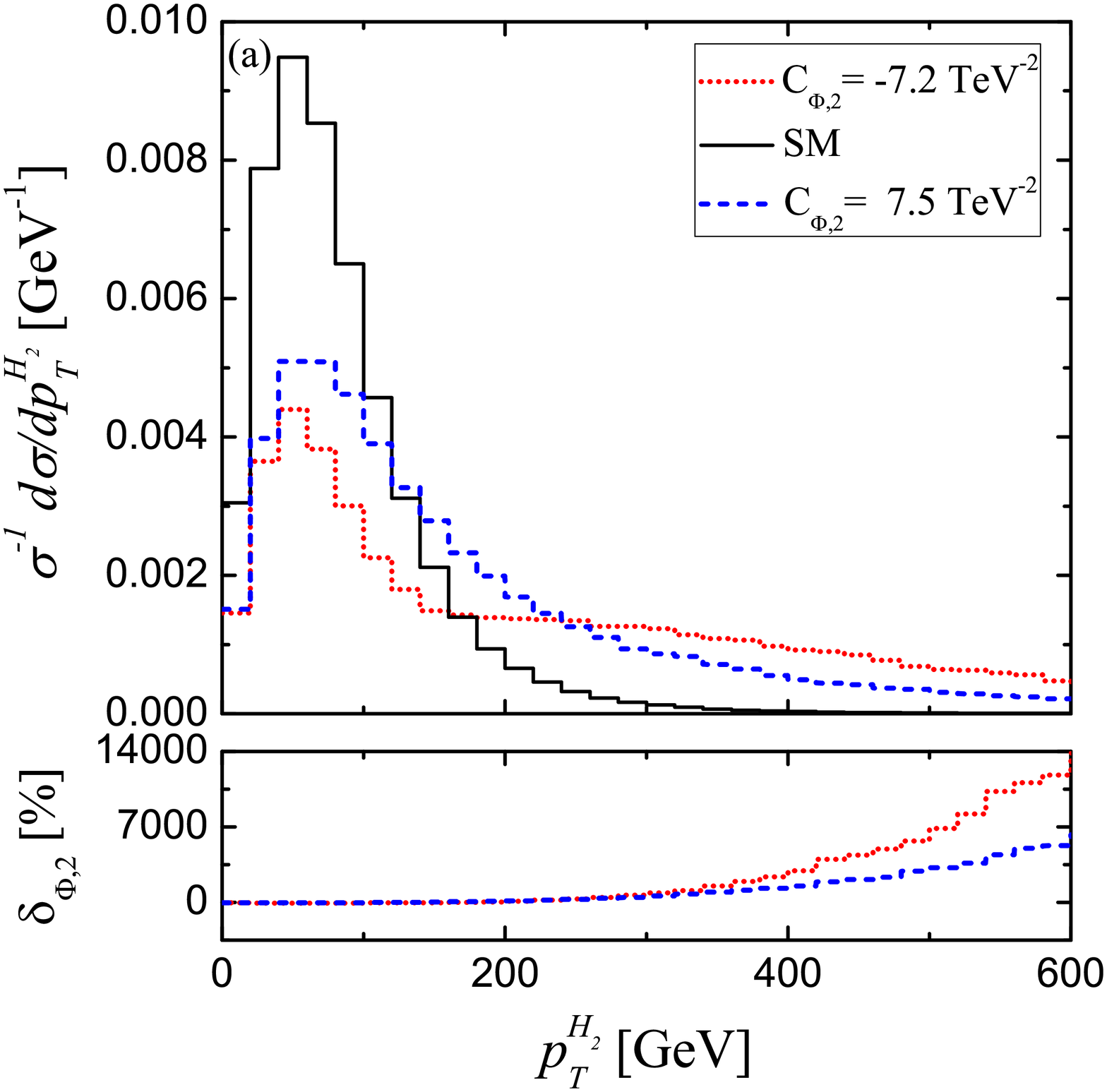}
\includegraphics[scale=0.29]{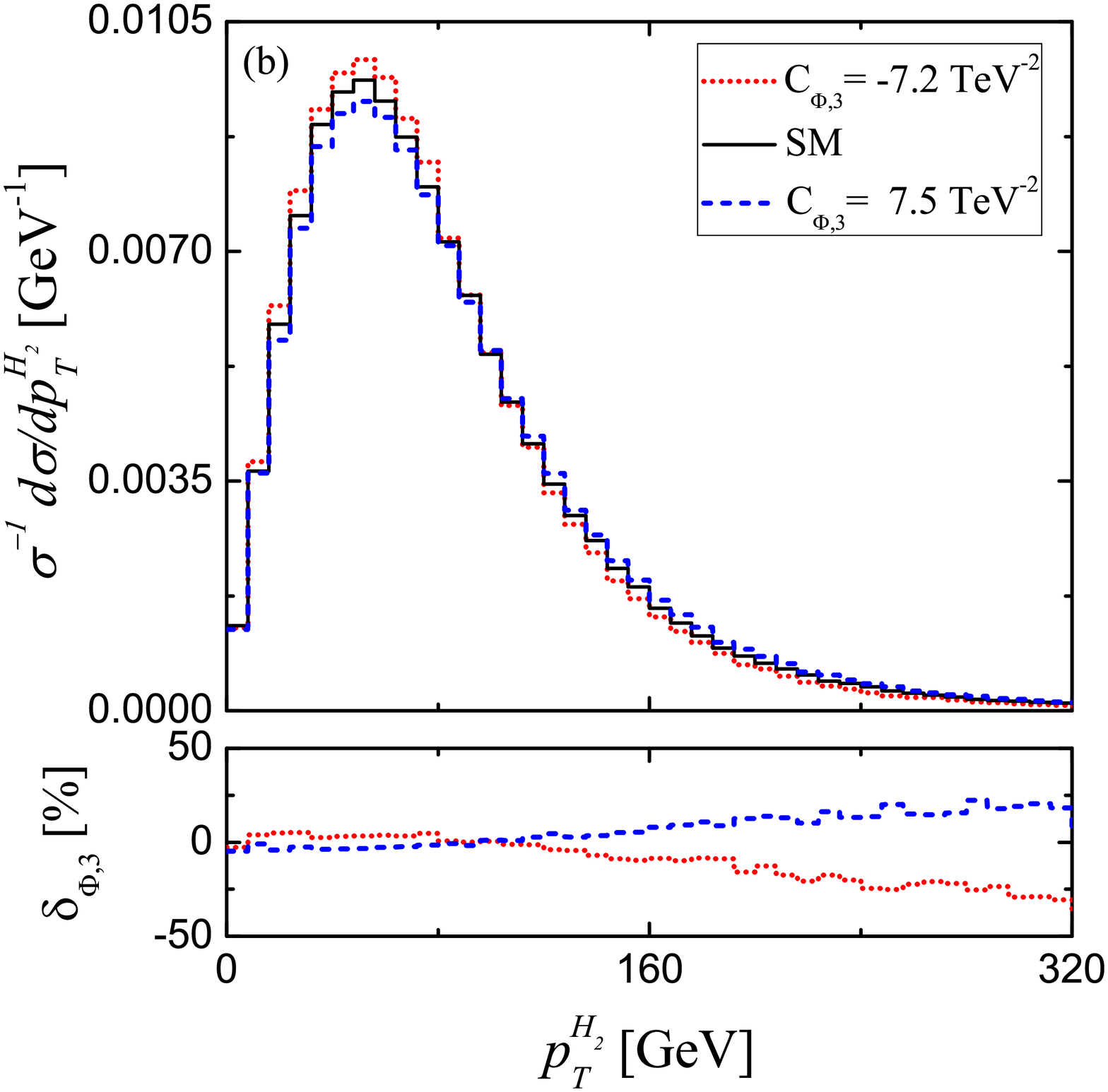}
\caption{
\label{fig7}
The same as Fig.\ref{fig5} but for the transverse momentum distributions of $H_2$. }
\end{center}
\end{figure*}

The two final-state Higgs bosons are named as the first Higgs boson $H_1$ and the second Higgs boson $H_2$ according to their transverse momenta, i.e., $p_T^{H_1} > p_T^{H_2}$. We present the normalized transverse momentum distributions and the corresponding relative discrepancies of the two Higgs bosons in Figs.\ref{fig6}(a, b) and \ref{fig7}(a, b), respectively. From Fig.\ref{fig6}(a) and Fig.\ref{fig7}(a) we see that the operator $\mathcal{O}_{\Phi,2}$ induces substantial changes on the shapes of the normalized transverse momentum distributions of both two Higgs bosons. The normalized $p_T^{H_1}$ $(p_T^{H_2})$ distributions for both $\mathcal{C}_{\Phi,2}=7.5~{\rm TeV^{-2}}$ and $\mathcal{C}_{\Phi,2}=-7.2~{\rm TeV^{-2}}$ are obviously enhanced in the region of $p_T^{H_1} > 300~ {\rm GeV}$ ($p_T^{H_2} > 200~ {\rm GeV}$) compared to the SM prediction, and the corresponding relative discrepancies are monotonically increasing functions of $p_T^{H_1}$ ($p_T^{H_2}$), particularly in high transverse momentum region.
Because the strength of triple Higgs self-interaction induced by operator ${\mathcal{O}}_{\Phi,2}$ is relevant to the momenta of the Higgs fields (see the $H(\partial_{\mu}H)(\partial^{\mu}H)$ term in Eq.(\ref{l-vertex-3})), the transverse momentum distributions of the two final Higgs bosons are apparently enhanced in high $p_T^{H_1}$ and $p_T^{H_2}$ regions from the corresponding SM predictions (see Fig.\ref{fig6}(a) and Fig.\ref{fig7}(a)). While Fig.\ref{fig6}(b) and Fig.\ref{fig7}(b) demonstrate that the impacts of the operator coefficients $\mathcal{C}_{\Phi,3}=7.5~{\rm TeV^{-2}}$ and $\mathcal{C}_{\Phi,3}=-7.2~{\rm TeV^{-2}}$ are small, they do not change the normalized SM $p_T$ distribution shapes of the two final-state Higgs bosons in a noticeable way, and the corresponding relative discrepancies are also relatively small and stable in the plotted $p_T$ region.

\par
\begin{figure*}
\begin{center}
\includegraphics[scale=0.29]{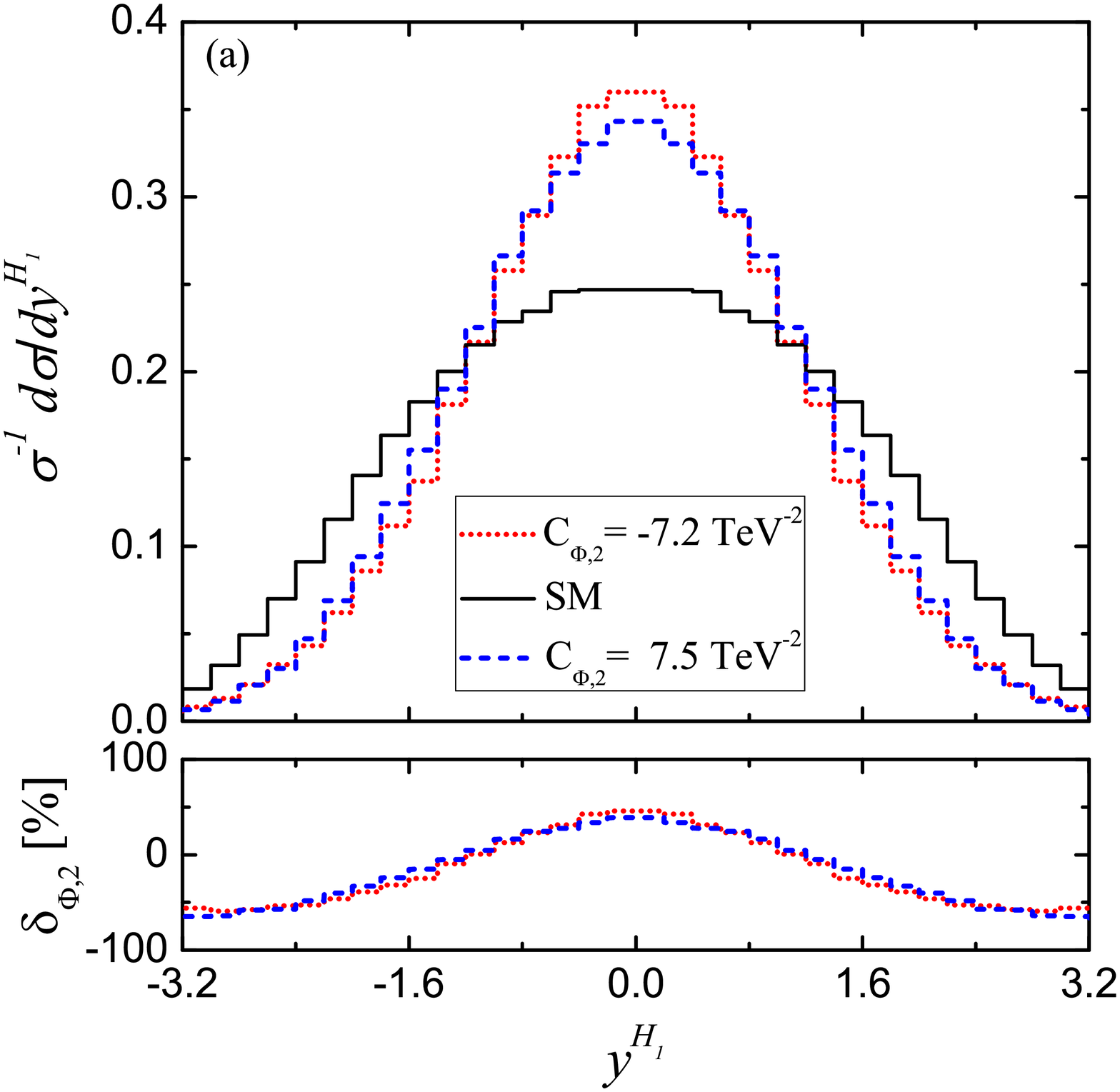}
\includegraphics[scale=0.29]{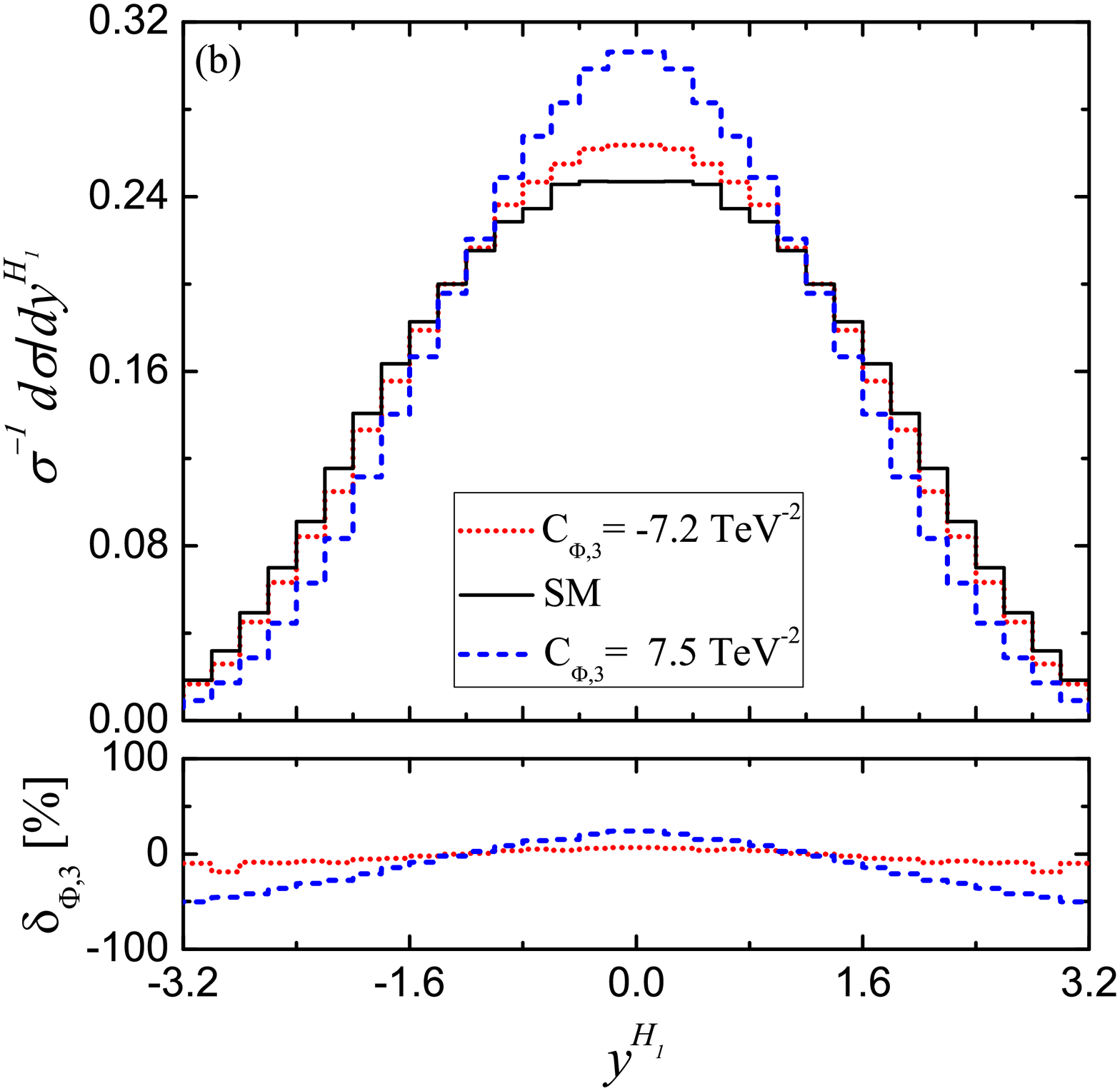}
\caption{
\label{fig8}
The same as Fig.\ref{fig5} but for the rapidity distributions of $H_1$.}
\end{center}
\end{figure*}
\begin{figure*}
\begin{center}
\includegraphics[scale=0.29]{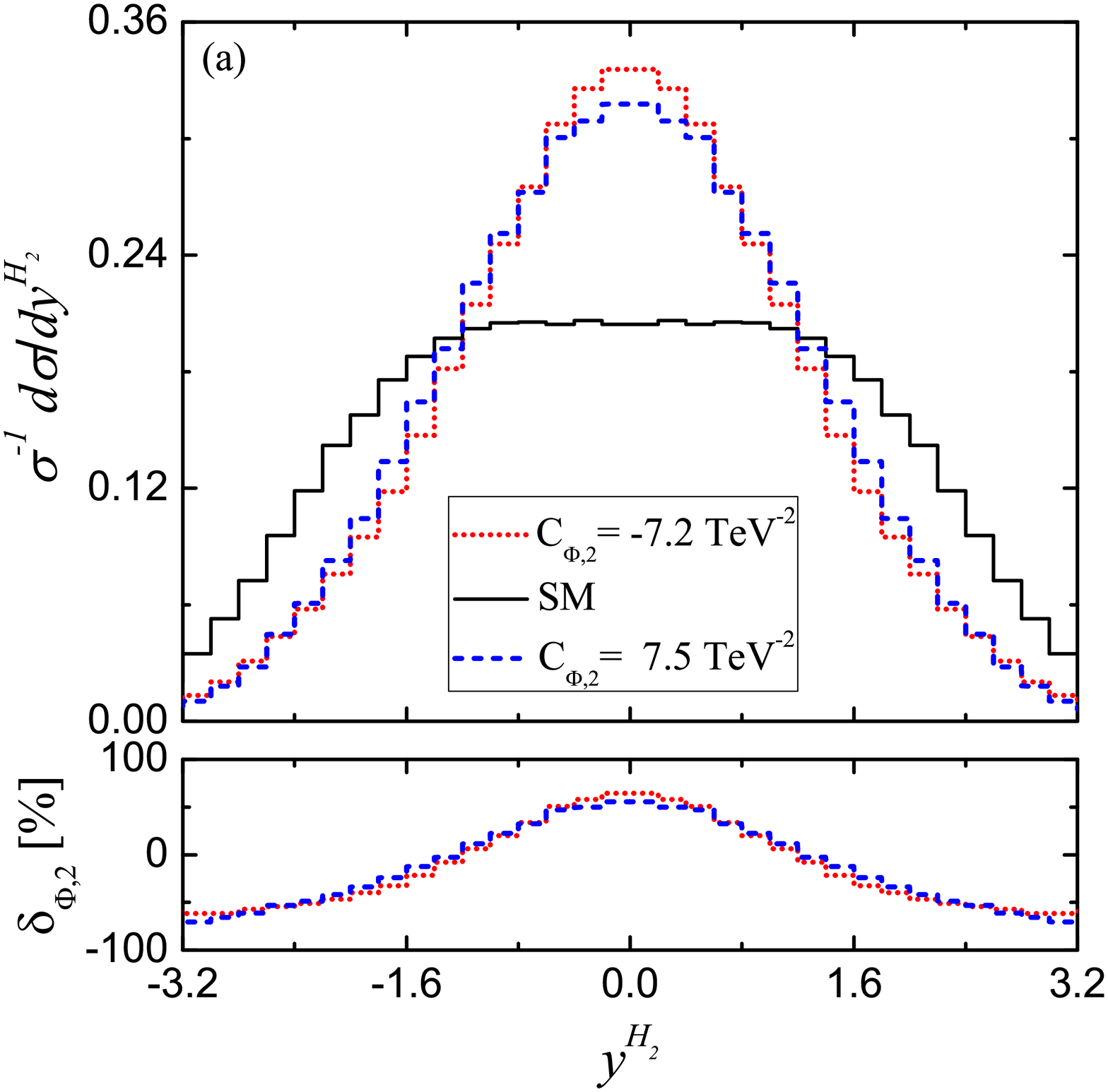}
\includegraphics[scale=0.29]{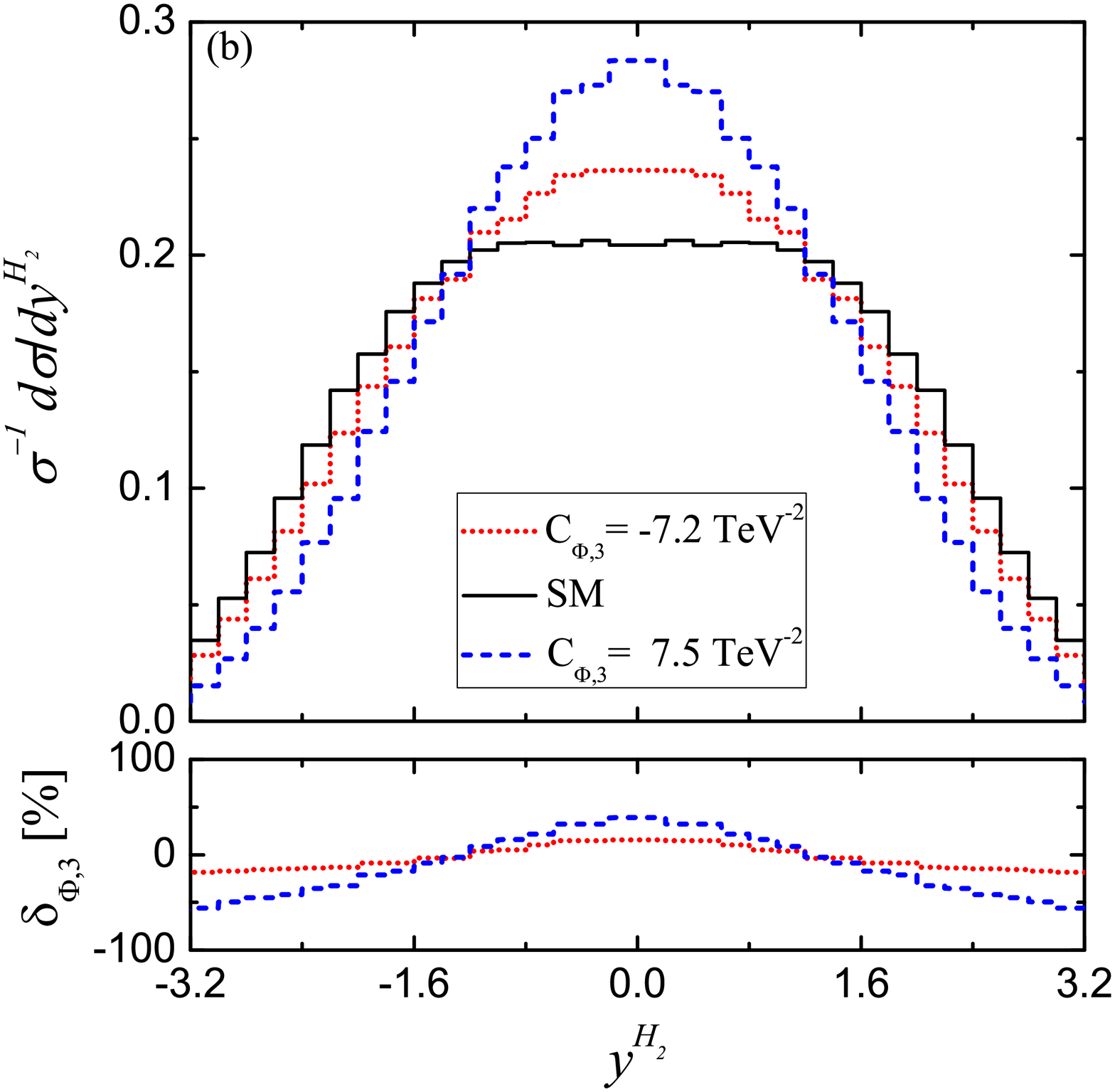}
\caption{
\label{fig9}
The same as Fig.\ref{fig5} but for the rapidity distributions of $H_2$.}
\end{center}
\end{figure*}

Figs.\ref{fig8}(a,b) and Figs.\ref{fig9}(a,b) provide the normalized rapidity distributions and the corresponding relative discrepancies of the two final-state Higgs bosons, separately. We find that each operator of $\mathcal{O}_{\Phi,2}$ and $\mathcal{O}_{\Phi,3}$ can induce noticeable changes on the shapes of the normalized rapidity distributions of the two Higgs bosons for the VBF Higgs pair production at the $14~{\rm TeV}$ LHC. The influences from these two operators on the normalized rapidity distributions of $H_1$ and $H_2$ are palpable, and alter the corresponding line-shapes of the distributions of $H_1$ and $H_2$ in a similar way. The relative discrepancies $\delta_{\Phi,i}(y^{H_{j}})~(i=2,3,~j=1,2)$ reach their maxima at $y^{H_j}=0$ and decrease slowly with the increment of $|y^{H_j}|$.

\par
\begin{figure*}
\begin{center}
\includegraphics[scale=0.29]{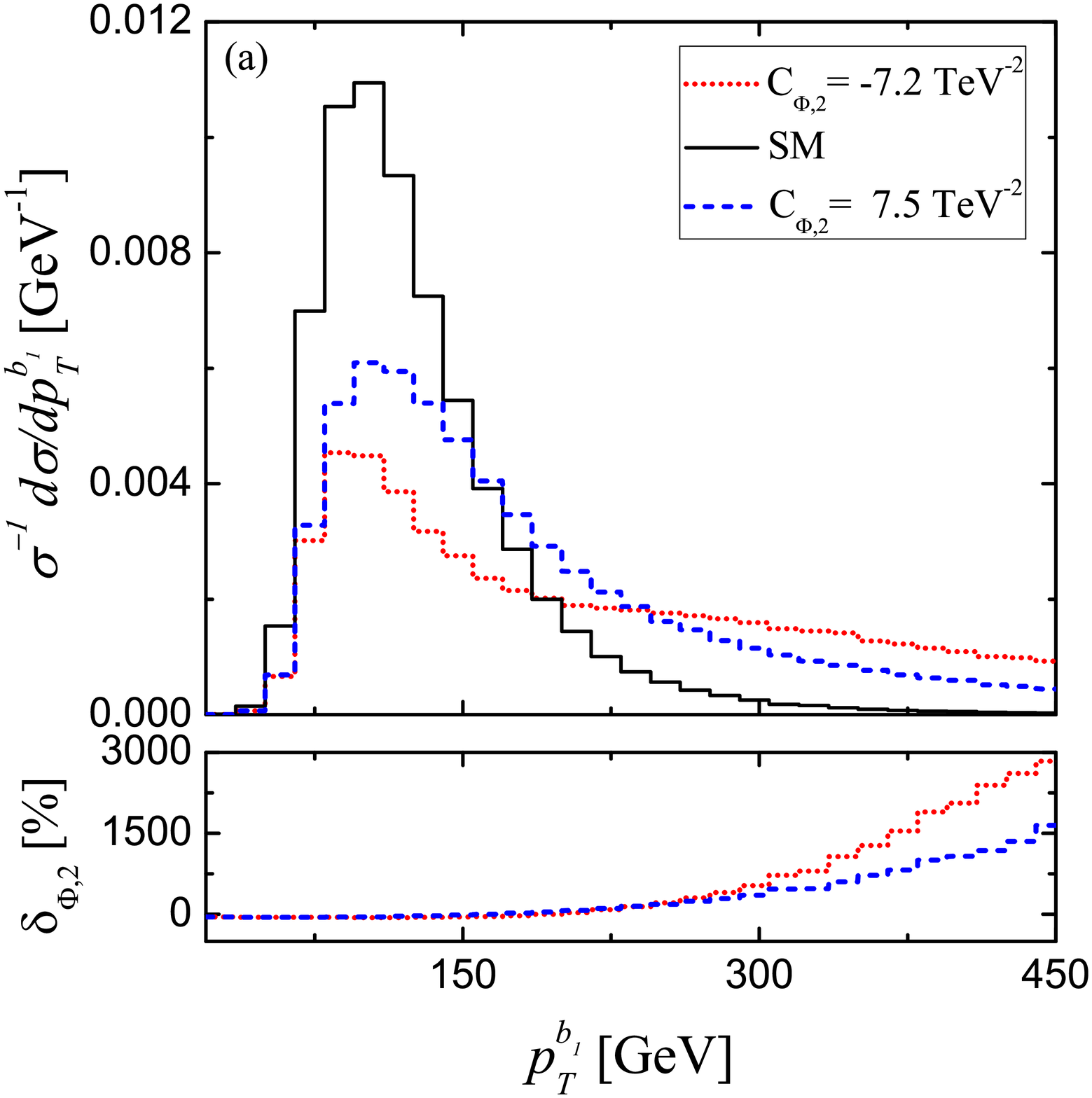}
\includegraphics[scale=0.29]{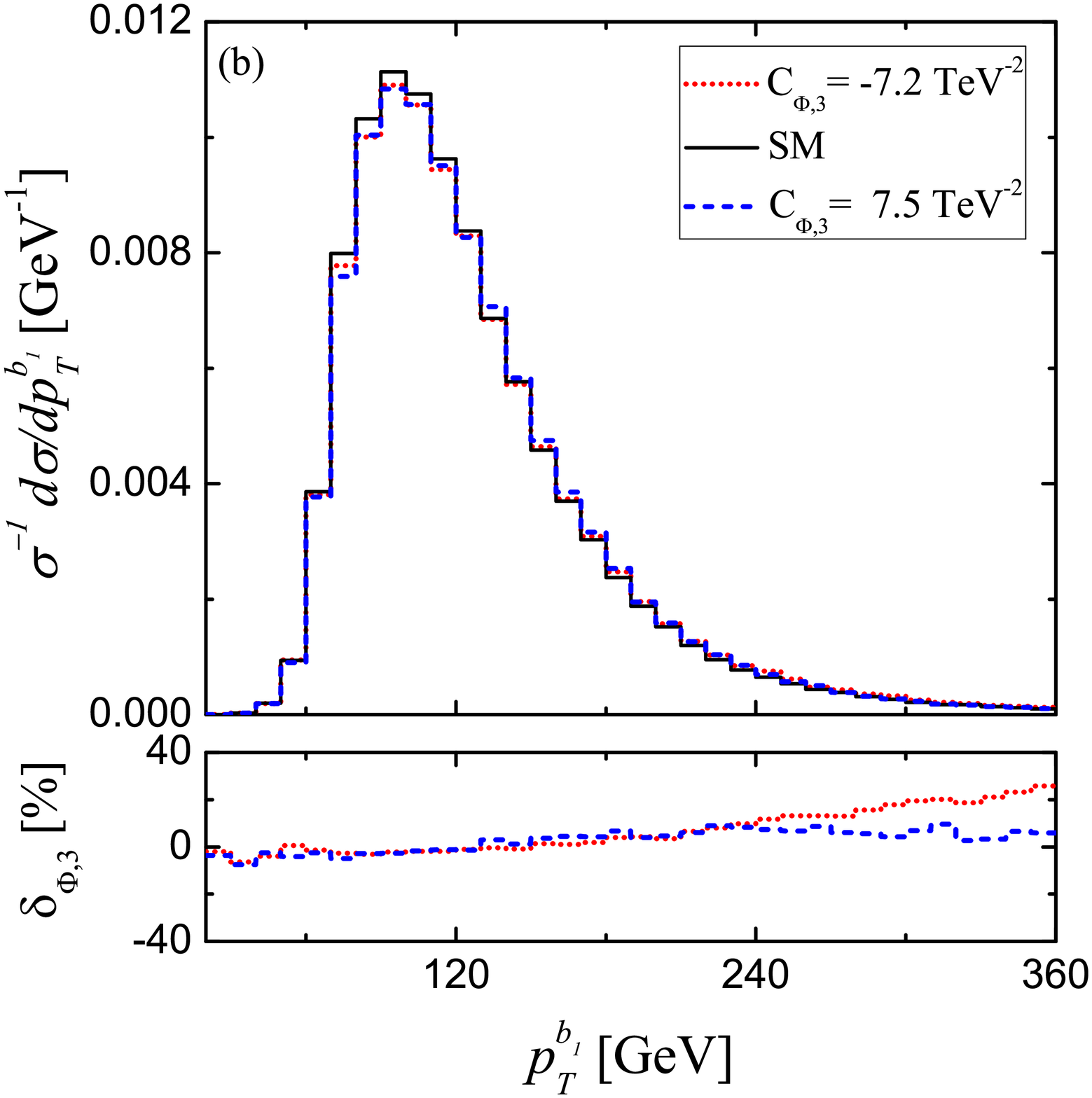}
\caption{ \label{fig10} The same as Fig.\ref{fig5} but for the transverse momentum distributions of the final leading b-jet $b_1$. }
\end{center}
\end{figure*}

We plot the normalized distributions of the leading b-jet transverse momentum $(p_T^{b_1})$ and the corresponding relative discrepancies $(\delta)$ with the constraints of $p_{T,b} \geq 20~{\rm GeV}$ and $|y_b| \leq 4.5$ for all the four final b-jets in Figs.\ref{fig10}(a) and (b). Fig.\ref{fig10}(a) provides the normalized $p_T^{b_1}$ distributions for the SM and the BSM cases with $\{\mathcal{C}_{BB} = \mathcal{C}_{WW} = \mathcal{C}_{\Phi,3} = 0,~\mathcal{C}_{\Phi,2}=-7.2~{\rm TeV^{-2}} \}$ and $\{ \mathcal{C}_{BB} = \mathcal{C}_{WW} = \mathcal{C}_{\Phi,3} = 0,~\mathcal{C}_{\Phi,2}=7.5~{\rm TeV^{-2}} \}$. In Fig.\ref{fig10}(b) the results are for the SM, $\{ \mathcal{C}_{BB} = \mathcal{C}_{WW} = \mathcal{C}_{\Phi,2} = 0,~\mathcal{C}_{\Phi,3}=-7.2~{\rm TeV^{-2}} \}$, and $\{ \mathcal{C}_{BB} = \mathcal{C}_{WW} = \mathcal{C}_{\Phi,2} = 0,~\mathcal{C}_{\Phi,3}=7.5~{\rm TeV^{-2}} \}$ separately. From these figures we see that the non-zero coefficient $\mathcal{C}_{\Phi,2}$ changes the SM distribution line shape clearly, the normalized $p_T^{b_1}$ distributions with non-zero $\mathcal{C}_{\Phi,2}$ are obviously enhanced in the region of $p_T^{b_1} > 200~{\rm GeV}$ in comparison with the SM prediction, while suppressed in the region of $p_T^{b_1} < 200~{\rm GeV}$. From Fig.\ref{fig10}(b) we see that the line-shape of the SM normalized $p_T^{b_1}$ distribution is not affected by the non-zero coefficient $\mathcal{C}_{\Phi,3}$ in a noticeable way.

\vskip 5mm
\section{Summary} \label{section-V}
\par
Higgs self-couplings trigger the EWSB and are very important to reveal the nature of the Higgs boson. We investigate the influences of the relevant dimension-six operators, especially $\mathcal{O}_{\Phi,2}$ and $\mathcal{O}_{\Phi,3}$, on the VBF Higgs boson pair production the at the $13$ and $14~ {\rm TeV}$ LHC. We included the NLO QCD corrections in our analysis for the purpose of reduce the remaining scale dependence. We find that the integrated cross section is particularly dependent on the dimension-six operators $\mathcal{O}_{\Phi,2}$ and $\mathcal{O}_{\Phi,3}$, and its measurement is suitable for ascertaining the limitation ranges of the coefficients $\mathcal{C}_{\Phi,2}$ and $\mathcal{C}_{\Phi,3}$. We systematically analyze the influences of these two dimension-six operators on the shapes of the normalized distributions of $M_{HH}$, $p_T^{H_1}$, $p_T^{H_2}$, $y^{H_1}$, $y^{H_2}$ and the leading b-jet $p_T^{b_1}$. We find that these kinematic distributions of the final Higgs bosons are affected by both $\mathcal{O}_{\Phi,2}$ and $\mathcal{O}_{\Phi,3}$ operators. While the normalized $p_T$ distribution of the leading b-jet is sensitive to operator $\mathcal{O}_{\Phi,2}$, but not to operator $\mathcal{O}_{\Phi,3}$. Particularly the $\mathcal{O}_{\Phi,2}$ can induce expressly quite different distribution shapes from the corresponding SM prediction. These features are very helpful in discriminating the signature of the new physics from the SM background, and distinguishing which dimension-six operator initiates the physics beyond the SM. The 5$\sigma$ discovery and 3$\sigma$ exclusion limits on the coefficients $\mathcal{C}_{\Phi,2}$ and $\mathcal{C}_{\Phi,3}$ by measuring the VBF $pp \to HH+X \to bb\bar{b}\bar{b}+X$ process at the $14~ {\rm TeV}$ LHC are also discussed.

\vskip 5mm
\par
\noindent{\large\bf Acknowledgments:}
This work was supported in part by the National Natural Science Foundation of China (No.11375171, No.11405173, No.11535002).


\vskip 5mm
\begin{thebibliography}{99}
\bibitem{hd-1}
G. Aad {\it et al.} (ATLAS Collaboration), Phys. Lett. B {\bf 716}, 1 (2012);
S. Chatrchyan {\it et al.} (CMS Collaboration), Phys. Lett. B {\bf 716}, 30 (2012).

\bibitem{hd-2}
G. Aad {\it et al.} (ATLAS Collaboration), Phys. Lett. B {\bf 726}, 120 (2013);
G. Aad {\it et al.} (ATLAS Collaboration), J. High Energy Phys. 01 (2015) 069;
V. Khachatryan {\it et al.} (CMS Collaboration), Eur. Phys. J. C {\bf 75}, 212 (2015);
V. Khachatryan {\it et al.} (CMS Collaboration), Phys. Rev. D {\bf 92}, 012004 (2015).

\bibitem{self-1}
T. Plehn and M. Rauch, Phys. Rev. D {\bf 72}, 053008 (2005).

\bibitem{collider-2}
https://fcc.web.cern.ch; http://cepc.ihep.ac.cn/;
M. L. Mangano {\it et al.}, CERN-TH-2016-112, arXiv:1607.01831.

\bibitem{self-2}
A. Djouadi, W. Kilian, M. Muhlleitner, and P. M. Zerwas, Eur. Phys. J. C {\bf 10}, 45 (1999).

\bibitem{self-3}
M. J. Dolan, C. Englert, and M. Spannowsky, J. High Energy Phys. 10 (2012) 112.

\bibitem{selfdect}
G. Aad {\it et al.} (ATLAS Collaboration), Phys. Rev. D {\bf 94}, 052002 (2016);
G. Aad {\it et al.} (ATLAS Collaboration), Phys. Rev. D {\bf 92}, 092004 (2015);
G. Aad {\it et al.} (ATLAS Collaboration), Eur. Phys. J. C {\bf 75}, 412 (2015);
V. Khachatryan {\it et al.} (CMS Collaboration), CERN-EP-2016-050, arXiv:1603.06896;
V. Khachatryan {\it et al.} (CMS Collaboration), Phys. Lett. B {\bf 749}, 560 (2015).

\bibitem{eft-1}
W. Buchmuller and D. Wyler, Nucl. Phys. B {\bf 268}, 621 (1986);
C. N. Leung, S. Love, and S. Rao, Z. Phys. C {\bf 31}, 433 (1986).

\bibitem{Henning}
B. Henning, X. C. Lu, and H. Murayama, J. High Energy Phys. 01 (1016) 023.

\bibitem{hpair-tot}
J. Baglio, A. Djouadi, R. Grober, M. M. Muhlleitner, J. Quevillon, and M. Spira, J. High Energy Phys. 04 (2013) 151.

\bibitem{hpair-1}
F. Goertz, A. Papaefstathiou, L. L. Yang, and J. Zurita, J. High Energy Phys. 04 (2015) 167.

\bibitem{hpair-2}
A. Azatov, R. Contino, G. Panico, and M. Son, Phys. Rev. D {\bf 92}, 035001 (2015).

\bibitem{hpair-3}
L. Edelhaeuser, A. Knochel, and T. Steeger, J. High Energy Phys. 11 (2015) 062.

\bibitem{hpair-4}
R. Grober, M. Muhlleitner, M. Spira, and J. Streicher, J. High Energy Phys. 09 (2015) 092.

\bibitem{hpair-5}
H. J. He, J. Ren, and W. M. Yao, Phys. Rev. D {\bf 93}, 015003 (2016).

\bibitem{hpair-6}
A. Carvalho, M. Dall'Osso, P. De Castro Manzano, T. Dorigo, F. Goertz, M. Gouzevich, and M. Tosi,
LHCHXSWG-2016-001, arXiv:1608.06578.

\bibitem{hpair-7}
C. R. Chen and I. Low, Phys. Rev. D {\bf 90}, 013018 (2014).

\bibitem{hpair-8}
C. T. Lu, J. Chang, K. Cheung, and J. S. Lee, J. High Energy Phys. 08 (2015) 133.

\bibitem{hpair-9}
D. Florian, I. Fabre, and J. Mazzitelli, ICAS-28/17, ZU-TH-08/17, arXiv:1704.05700.

\bibitem{vbfsig}
A. Djouadi, Phys. Rep. {\bf 457}, 1 (2008).

\bibitem{eft-2}
B. Grzadkowski, M. Iskrzynski, M. Misiak, and J. Rosiek, J. High Energy Phys. 10 (2010) 085;
K. Hagiwara, S. Ishihara, R. Szalapski, and D. Zeppenfeld, Phys. Rev. D {\bf 48}, 2182 (1993).

\bibitem{Andre}
A. de Gouvea, J. Herrero-Garcia, and A. Kobach, Phys. Rev. D {\bf 90}, 016011 (2014);
A. Kobach, Phys. Lett. B {\bf 758}, 455 (2016).

\bibitem{Ben}
B. Gripaios, arXiv:1503.02636.

\bibitem{Grogean}
C. Grojean, E. E. Jenkins, A. V. Manohar, and M. Trott, J. High Energy Phys. 04 (2013) 016;
J. Elias-Miro, J. R. Espinosa, E. Masso, and A. Pomarol, J. High Energy Phys. 08 (2013) 033;
J. Elias-Miro, J. R. Espinosa, E. Masso, and A. Pomarol, J. High Energy Phys. 11 (2013) 066;
E. E. Jenkins, A. V. Manohar, and Michael Trott,J. High Energy Phys. 10 (2013) 087;
E. E. Jenkins, A. V. Manohar, and M. Trott, J. High Energy Phys. 01 (2014) 035;
R. Alonso, E. E. Jenkins, A. V. Manohar, and M. Trott, J. High Energy Phys. 04 (2014) 159;
J. Elias-Miro, C. Grojean, R. S. Gupta, and D. Marzocca, J. High Energy Phys. 05 (2014) 019;
R. Alonso, H. M. Chang, E. E. Jenkins, A. V. Manohar, and B. Shotwell, Phys. Lett. B {\bf 734}, 302 (2014).

\bibitem{eft-5}
T. Corbett, O. J. P. Eboli, and M. C. Gonzalez-Garcia, Phys. Rev. D {\bf 91}, 035014 (2015).

\bibitem{Roberto}
R. Contino, A. Falkowski, F. Goertz, C. Grojean, and F. Riva, J. High Energy Phys. 07 (2016) 144.

\bibitem{eft-3}
T. Corbett, O. J. P. Eboli, J. Gonzalez-Fraile, and M. C. Gonzalez-Garcia, Phys. Rev. D {\bf 87}, 015022 (2013).

\bibitem{constraint-1}
A. Falkowski and F. Riva, J. High Energy Phys. 02 (2015) 039.

\bibitem{st-1}
K. Hagiwara, R. D. Peccei, D. Zeppenfeld, and K. Hikasa, Nucl. Phys. B {\bf 282}, 253 (1987);
M. E. Peskin and T. Takeuchi, Phys. Rev. Lett. {\bf 65}, 964 (1990);
S. Alam, S. Dawson, and R. Szalapski, Phys. Rev. D {\bf 57}, 1577 (1998).

\bibitem{eft-4}
M. Baak, J. Cuth, J. Haller, A. Hoecker, R. Kogler, K. Moenig, M. Schott, and J. Stelzer,
Eur. Phys. J. C {\bf 74}, 3046 (2014);
J. Ellis, V. Sanz, and T. You, J. High Energy Phys. 03 (2015) 157.

\bibitem{vbfnnlo}
L. S. Ling, R. Y. Zhang, W. G. Ma, L. Guo, W. H. Li, and X. Z. Li, Phys. Rev. D {\bf 89}, 073001 (2014).

\bibitem{program-2}
J. Alwall, R. Frederix, S. Frixione, V. Hirschi, F. Maltoni, O. Mattelaer, H. S. Shao, T. Stelzer, P. Torrielli,
and M. Zaro, J. High Energy Phys. 07 (2014) 079.

\bibitem{program-3}
A. Alloul, N. D. Christensen, C. Degrande, C. Duhr, and B. Fuks, Comput. Phys. Commun. {\bf 185}, 2250 (2014);
A. Alloul, B. Fuks, and V. Sanz, J. High Energy Phys. 04 (2014) 110.

\bibitem{vbfcut1}
T. Figy and D. Zeppenfeld, Phys. Lett. B {\bf 591} 297 (2004);
T. Figy, Mod. Phys. Lett. A {\bf 23}, 1961 (2008);
M. Ciccolini, A. Denner, and S. Dittmaier, Phys. Rev. Lett. {\bf 99}, 161803 (2007).

\bibitem{vbfcut2}
N. Greiner, S. Hoeche, G. Luisoni, M. Schonherr, J. C. Winter, and V. Yundin, J. High Energy Phys. 01 (2016) 169;
N. Greiner, S. Hoeche, G. Luisoni, M. Schonherr, J. C. Winter, and V. Yundin, MSUHEP-160114,
SLAC-PUB-16455, ZU-TH-1/16, arXiv:1601.03722.

\bibitem{mstw}
A. D. Martin, W. J. Stirling, R. S. Thorne, and G. Watt, Eur. Phys. J. C {\bf 63}, 189 (2009).

\bibitem{pdg}
K. A. Olive {\it et al.} (Particle Data Group), Chin. Phys. C {\bf 38},
090001 (2014) and 2015 update.

\bibitem{scale_choice}
M. Cacciari, F. A. Dreyer, A. Karlberg, G. P. Salam, and G. Zanderighi, Phys. Rev. Lett. 115, 082002 (2015).

\bibitem{sfa}
T. Han, G. Valencia, and S. Willenbrock, Phys. Rev. Lett. {\bf 69}, 3274 (1992).

\bibitem{constraint}
A. Butter, O. J. P. Eboli, J. Gonzalez-Fraile, M. C. Gonzalez-Garcia, T. Plehn, and M. Rauch,
J. High Energy Phys. 07 (2016) 152.

\bibitem{VBFNLO}
J. Baglio {\it et al.}, CERN-PH-TH/2011-173, DESY 11-125, arXiv:1107.4038;
J. Baglio {\it et al.}, FTUV-14-2903, IFIC-14-26, arXiv:1404.3940.

\bibitem{hdecay}
A. Djouadi, J. Kalinowski, and M. Spira, Comput. Phys. Commun. {\bf 108}, 56 (1998).

\end{thebibliography}
\end{document}